\documentclass[12pt,preprint]{aastex}

\newcommand{\myemail}{amonreal@aip.de}
\newcommand{\gsim}{\hbox{\rlap{\lower.55ex\hbox{$\sim$}} \kern-.3em
\raise.4ex \hbox{$>$}}}
\newcommand{\lsim}{\hbox{\rlap{\lower.55ex\hbox{$\sim$}} \kern-.3em
\raise.4ex \hbox{$<$}}}

\slugcomment{to be submitted to The ApJ
             (Version 1.0 13/09/2005 ami)}

\shorttitle{LINER spectra in ULIRGs}
\shortauthors{Monreal-Ibero et al.}

\begin{document}

\title{LINER-like Extended Nebulae in ULIRGs: Shocks Generated by Merger
  Induced Flows} 

\author{A. Monreal-Ibero\altaffilmark{1}} 
\affil{Instituto de Astrof\'{\i}sica de Canarias, c/ V\'{\i}a L\'actea s/n,
  38200 La Laguna}
\email{\myemail}

\author{S. Arribas\altaffilmark{2,3}}
\affil{Space Telescope Science Institute, Baltimore, MD 21218}
\email{arribas@stsci.edu}

\and

\author{L. Colina}
\affil{Instituto de Estructura de la Materia (CSIC), Serrano 121, 28006 Madrid}
\email{colina@isis.iem.csic.es}

\begin{abstract}

In this work we studied the two-dimensional ionization structure of the
circumnuclear and extranuclear regions in a sample of six low$-z$
Ultraluminous Infrared Galaxies using Integral Field Spectroscopy.  
The ionization conditions in the extranuclear regions of these galaxies ($\sim
5 - 15$~kpc) are typical of LINERs as obtained from the Veilleux-Osterbrock
line ratio diagnostic diagrams. The range of observed line ratios is best
explained by the presence of fast shocks with velocities of 150 to 500 
km s$^{-1}$, while the ionization by an AGN or nuclear starburst is in general
less likely. The comparison of the two-dimensional ionization level and
velocity dispersion in the extranuclear regions of these galaxies shows a
positive correlation, further supporting the idea that shocks are indeed the
main cause of ionization.  

The origin of these shocks is also investigated. Despite the likely presence
of superwinds in the  circumnuclear regions of these systems, no evidence for
signatures of superwinds such as double velocity components  are found in the 
extended extranuclear regions. We consider a more likely explanation for the
presence of shocks, the  existence of tidally induced large scale gas flows
caused by the merging process itself, as evidenced by the observed 
velocity fields characterized by peak-to-peak velocities of 400 km s$^{-1}$,
and velocity dispersions  of up  to 200 km s$^{-1}$.  

\end{abstract}


\altaffiltext{1}{Now in Astrophysikalisches Institut Potsdam, An der
  Sternwarte, 16, D~14482 Potsdam } 
\altaffiltext{2}{On leave from the Instituto de Astrof\'{\i}sica de Canarias
  (CSIC)}
\altaffiltext{3}{Affiliated with the Space Telescope Division of the European 
Space Agency,  ESTEC, Nordwijk, Netherlands}
 
\section{INTRODUCTION}

Ultraluminous Infrared Galaxies (ULIRGs), defined as objects with an
infrared luminosity similar to that of optically selected quasars
($L_{bol} \approx L_{IR} \ga 10^{12} L_{\sun}$) may be the local counterpart
of some high-z galaxy populations \citep[see][]{san96,gen00,fra03,lef04}. Most
(if not all) of them show signs of mergers and interactions
\citep[e.g.][]{cle96,sco00,sur00b,bor00} and it has been found that they could
be the progenitors of intermediate-mass elliptical galaxies \citep[][and
references therein]{gen01,tac02}. They have large amounts of gas and dust and 
they are undergoing intense starburst activity. Some of these objects may
harbor an active galacic nucleus (AGN) although its importance as source of
energy in ULIRGs is still under debate.   

The origin of the ionization of gas in these objects has been mainly studied
in the innermost (nuclear) regions \citep[e.g.][]{kim98}.  These studies show
that $\sim 35 \%$ of their nuclei have a LINER-like ionization independently  
of the luminosity while the fraction of Seyfert-like spectrum increases with
luminosity. However, due to their complex structure, studies based on nuclear
optical spectroscopy may lead to misclassifications. This can be due to
several causes. For instance, the actual nucleus of the system may be obscured
in the optical or, alternatively, the dominant region in the emission-line may
not be coincident  with the nucleus. An example where both effects have been
reported is IRAS~12112+0305 \citep{col00}.  In addition, standard slit
spectroscopic observations may be affected by other type of technical
uncertainties like, for instance, misalignment of the slit, differential
atmospheric refraction, etc. 

ULIRGs, as systems that are undergoing an intense starburst activity phase (and
with AGN activity in some cases), are good candidates to produce superwinds
\citep[see][]{vei05}. Evidence of superwinds has already been reported in
several systems, on the basis of the properties of the 
emission \citep{hec90,leh96,arr01} and absorption
\citep{hec00,rup02,rup05a,rup05b,mar05} lines. Although superwinds are
probably playing a role in the ionization of the circumnuclear region of
ULIRGs, their importance in the extranuclear regions is unclear. In them,
tidally induced forces associated to the interaction process  
itself has been also proposed as the mechanism responsible of shocks
\citep{mcd03,col04}.

The present article is focused in the study of the 2D structure and mechanisms
of ionization of six ULIRGs based on the use of Integral Field Spectroscopy
(IFS). This observational technique is well suited for this goal since it
allows to obtain simultaneously spectral information of in a 2D field. The
present work is part of a program aimed at studying the internal structure and
kinematics of (U)LIRGs, on the basis of this technique and high-resolution
images obtained with HST  \citep[see][and refereces therein]{col05}.  The
paper is structured as follows: In section \S2, we briefly describe the sample
of galaxies analysed and summarize how observations where performed. In
section \S3, the reduction process and data analysis are described. Section
\S4 presents the results obtained both in the external parts and the nuclear
regions and discuss the mechanisms responsible of the observed ionization.
Section \S5 summarize the main conclusions.

Thoroughout the paper, a Hubble constant of 70~km~s$^{-1}$~Mpc$^{-1}$ is
assumed. This implies a linear scale between 0.89  and 2.58~kpc arcsec$^{-1}$
for the systems analyzed.

\section{SAMPLE AND OBSERVATIONS} 

\subsection{The Sample of Galaxies}

The sample of galaxies consists of six low-{\it z} ULIRGs (see 
general properties in table \ref{misulirgs}) covering
a relative wide range of dynamical states of the merging process.   
Three of the galaxies (IRAS~08572+3915, IRAS~12112+0305, and
IRAS~14348$-$1447) are interacting pairs separated by projected distances of up
to 6 kpc, while the rest of the galaxies (IRAS~15206+3342, IRAS~15250+3609,
and  IRAS~17208$-$0014) are more evolved, single nucleus ULIRGs, some with
a light profile and overall kinematics close to that of intermediate
mass ellipticals (e.g. IRAS~17208$-$0014, Genzel et al. 2001).
The two-dimensional kinematic properties (velocity field and velocity
dispersion) have been studied in detail before using integral field
optical spectroscopy (see Colina et al. 2005 and references therein).

The complexity of the two-dimensional ionization field in these galaxies
is such that previous long-slit spectroscopic studies have classified
the nucleus of several of the galaxies differently. For example,
IRAS~08572+3915 was originally classified as Seyfert 2 \citep{san88} although
both nuclei were classified as LINERs later on \citep{kim98};
IRAS~14348$-$1447 has been classified either as LINER \citep{kim98} or 
Seyfert 2 \citep{san88}; IRAS~15206+3342 has been identified both as a
Seyfert 2 \citep{san88,sur00a} and as an \textsc{H~ii} 
\citep{kim98}, and IRAS~15250+3609 is classified both as \textsc{H~ii} 
and LINER \citep{kim95,baa98}. The other two galaxies, IRAS~12112+0305
and IRAS~17208$-$0014 are classified as LINER
\citep{vei99} and \textsc{H~ii} \citep{kim95}, respectively. Our IFS data
disagree with some of the previous classifications, in particular the two
nuclei of IRAS~08572+3915 are classified as \textsc{H~ii} (Arribas et
al. 2000), and the true, optically hidden, nucleus of IRAS~17208$-$0014 is
classified as a LINER \citep{arr03}. 

\subsection {Observations}

Data were obtained with the INTEGRAL system \citep{arr98} plus the WYFFOS
spectrograph \citep{bin94} in the 4.2~m WHT at the Observatorio del Roque de
los Muchachos (Canary Islands). Spectra were taken using the fiber bundle SB2
and a 600 lines mm$^{-1}$ grating with an effective resolution of 
$\sim$4.8~\AA. Fibers in an INTEGRAL bundle are arranged in two sets: most of
them (189 for SB2) form a rectangular area centered on the
object while the rest of them form a circle around it and observe
simultaneously the sky. In the case of SB2 the covered field is of
16\farcs5$\times$12\farcs3. 
Data were taken under photometric conditions and the seeing was of $\sim$
1\farcs0--1\farcs5 except for the 1 April 1998 observing run when it was about
2\farcs0. Table \ref{obs} summarises the parameters of the
observations. Besides, HST imaging in the I-band (WFPC2
F814W filter) is available for all of
them and, with the exception  of \object{IRAS~15206+3342}, also in the 
H-band (NICMOS F160W filter).

\section{DATA REDUCTION AND ANALYSIS \label{reduc}}

The basic reduction process includes bias subtraction, scattered light
removal, extraction of the apertures, wavelength calibration, throughput and
flatfield correction, sky substraction, cosmic ray rejection and relative flux
calibration. Though it is not strictly necessary for the present paper, an
absolute flux calibration was also performed \citep{mon04}.

For the present analysis the strongest optical emission lines
including [\textsc{O~i}]$\lambda6300$, H$\alpha$,
[\textsc{N~ii}]$\lambda\lambda6548,6584$ and 
[\textsc{S~ii}]$\lambda\lambda6717,6730$ have been used. Each
emission line profile was fitted to a single Gaussian function using the
DIPSO package inside the STARLINK
environment\footnote{http://www.starlink.rl.ac.uk/}. The set of lines
H$\alpha$+[\textsc{N~ii}]$\lambda\lambda$6548,6584 was fitted simultaneously,
fixing the separation in wavelength between the three lines, assuming that all
lines had the same width and fixing the ratio between the nitrogen lines to
3. Sulfur lines were fitted fixing the distance between them and assuming the
same width for both lines. In all cases, a constant value was assigned to the
local continuum. A single (gaussian) component is in general
a good representation of the observed line profiles, with the exception
of some nuclear regions. For these regions a two-component fit was
necessary. The ionized gas velocity dispersion was derived from the
H$\alpha$ line width (after subtracting the instrumental profile in
quadrature). 

To study the ionization state, the line ratios
[\textsc{O~i}]$\lambda$6300/H$\alpha$, 
[\textsc{N~ii}]$\lambda$6584/H$\alpha$ 
and [\textsc{S~ii}]$\lambda\lambda$6717,6731/H$\alpha$ were
calculated for each spectrum. Since the H$\beta$ emission line was 
detected in an area substantially smaller than for the H$\alpha$, these line
ratios were not corrected from extinction. For the
[\textsc{N~ii}]$\lambda$6584/H$\alpha$ ratio, the two lines involved are so
close one from the other that reddening is negligible. In the case of the
[\textsc{S~ii}]$\lambda\lambda$6717,6731/H$\alpha$ ratio, if extinction is
moderate, the value of the ratio may change slightly, while in the regions
where it is more elevated ($E(B-V) \gsim 1.0$), the extinction produces
somewhat smaller ratios (typically by \lsim 0.1~dex). However, this small
difference does not change the main conclusions of the present analysis. To
better visualize the spatial distribution of the relevant magnitudes
(e.g. H$\alpha$ flux, velocity dispersion,
[\textsc{N~ii}]$\lambda$6584/H$\alpha$), two-dimensional images (maps) were
created using a Renka \& Cline two-dimensional interpolation method
(Fig. 1). All theses images have 81$\times$81  pixels, with an scale of
0\farcs21~pix$^{-1}$.

\section{RESULTS AND DISCUSSION}

The kinematical properties of the galaxies under analysis have already been
studied by \citet{col05}, who conclude that the global motions of the gas
(i.e. velocity fields) are dominated by merger-induced flows, showing
peak-to-peak velocity differences of $\sim$ 400~km~s$^{-1}$.  
Only one out of our six systems (\object{IRAS~17208$-$0014}) shows clear
evidences of ordered rotational motions although there are also some hints of
rotation in \object{IRAS~08572+3915}.  The ionized gas velocity
dispersion maps revealed high-velocity regions ($\sim$70-200~km~s$^{-1}$) that
do not trace any special mass concentration. 

\begin{figure}[htb!]
\centering
\epsscale{1.00}
\includegraphics[angle=0,scale=.9, clip=,bbllx=44, bblly=176, bburx=447,
bbury=662]{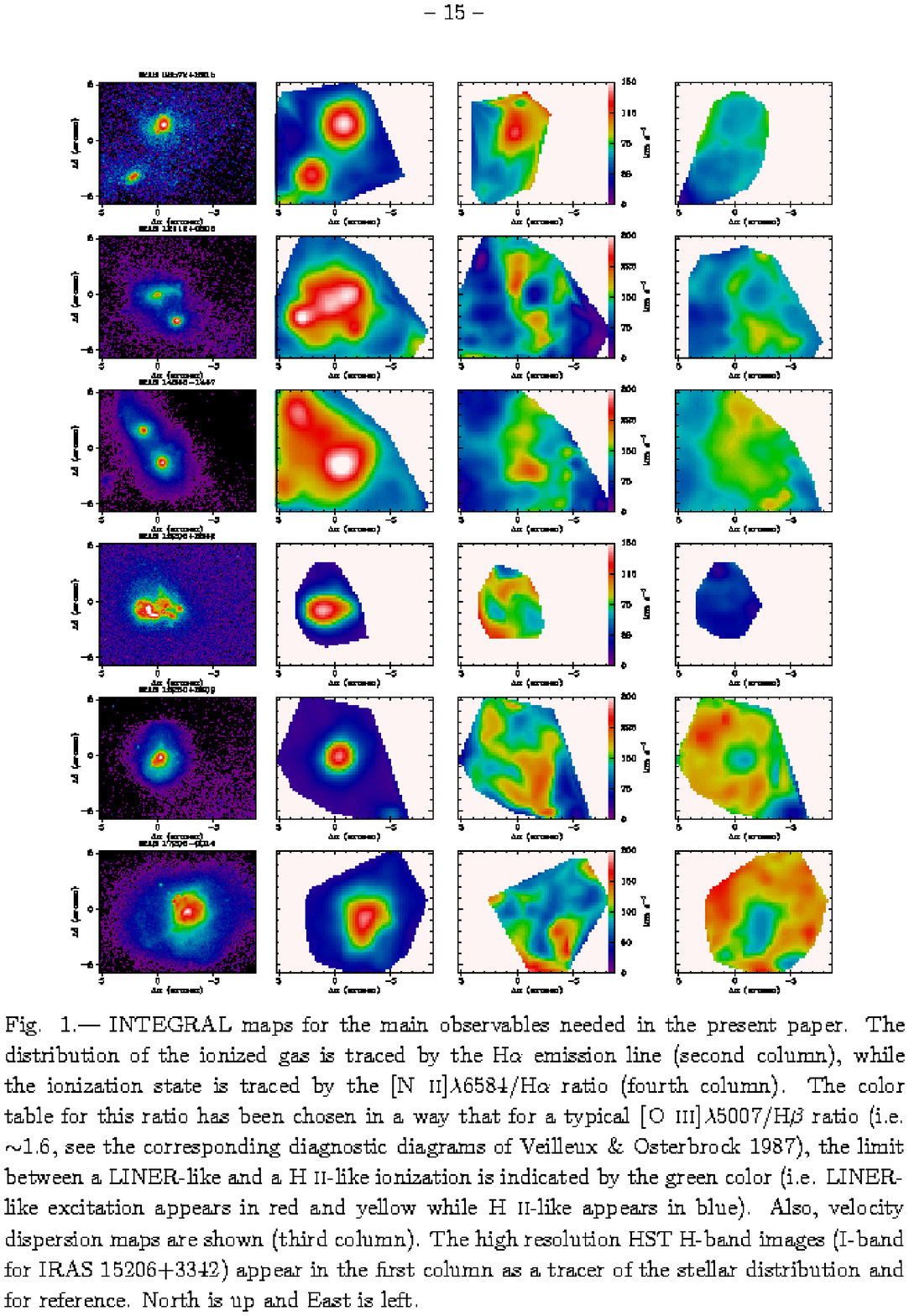}\vspace{2mm}
\protect\caption{INTEGRAL maps for the main observables needed in 
  the present paper. The distribution of the ionized gas is traced by the
  H$\alpha$ emission line (second column), while the ionization state is
  traced by the \textsc{[N$\;$ii]}$\lambda$6584/H$\alpha$ ratio (fourth
  column).  The color table for this ratio has been chosen in a way that for a
  typical 
\textsc{[O$\;$iii]}$\lambda$5007/H$\beta$ ratio \citep[i.e. $\sim$1.6, see 
the corresponding diagnostic diagrams of][]{vei87}, the limit
between a LINER-like and a \textsc{H$\;$ii}-like ionization is indicated by the
green color (i.e.  LINER-like excitation appears in red and yellow while
  \textsc{H$\;$ii}-like appears in blue). Also, velocity dispersion maps are
  shown (third column). The high resolution HST H-band images (I-band for
  IRAS~15206+3342) appear in the first column as a tracer of the stellar
  distribution and for reference. North is up and East is left.\label{panel}} 
\end{figure}

In the following section, we discuss the results derived from
Figure \ref{panel}, and in particular those from the H$\alpha$ and velocity
dispersion maps. 

\subsection{Two-dimensional Ionization Structure of the extranuclear Emission
  Line Nebulae}  

Typical values of  \textsc{[O$\;$iii]}$\lambda$5007/H$\beta$ 
in the brightest regions of these galaxies are around 1--2.5. Assuming similar
values for fainter regions (where this ratio cannot be obtained due to the
faintness of the lines), the  \textsc{[N$\;$ii]}$\lambda$6584/H$\alpha$ ratio
can be used to distinguish between LINER and  H~\textsc{ii}-like ionization 
\citep[see the diagnostic diagrams of ][]{vei87}. 

In general, the \textsc{[N$\;$ii]}$\lambda$6584/H$\alpha$ maps (see Fig. 1)
show a complex ionization structure.  According to this ratio, LINER-like
emission is found in the extended extranuclear regions in three systems: 
\object{IRAS~14348$-$1447}, \object{IRAS~15250+3609}, and  
\object{IRAS~17208$-$0014}. Contrary, for \object{IRAS~08572+3915},
\object{IRAS~15206+3342} and  \object{IRAS~12112+0305} this 
line ratio suggests a dominant H~\textsc{ii}-like ionization. 

\begin{figure}[htb!]
\centering
\includegraphics[width=13cm,angle=0, clip=,bbllx=30, bblly=210, bburx=560,
bbury=640]{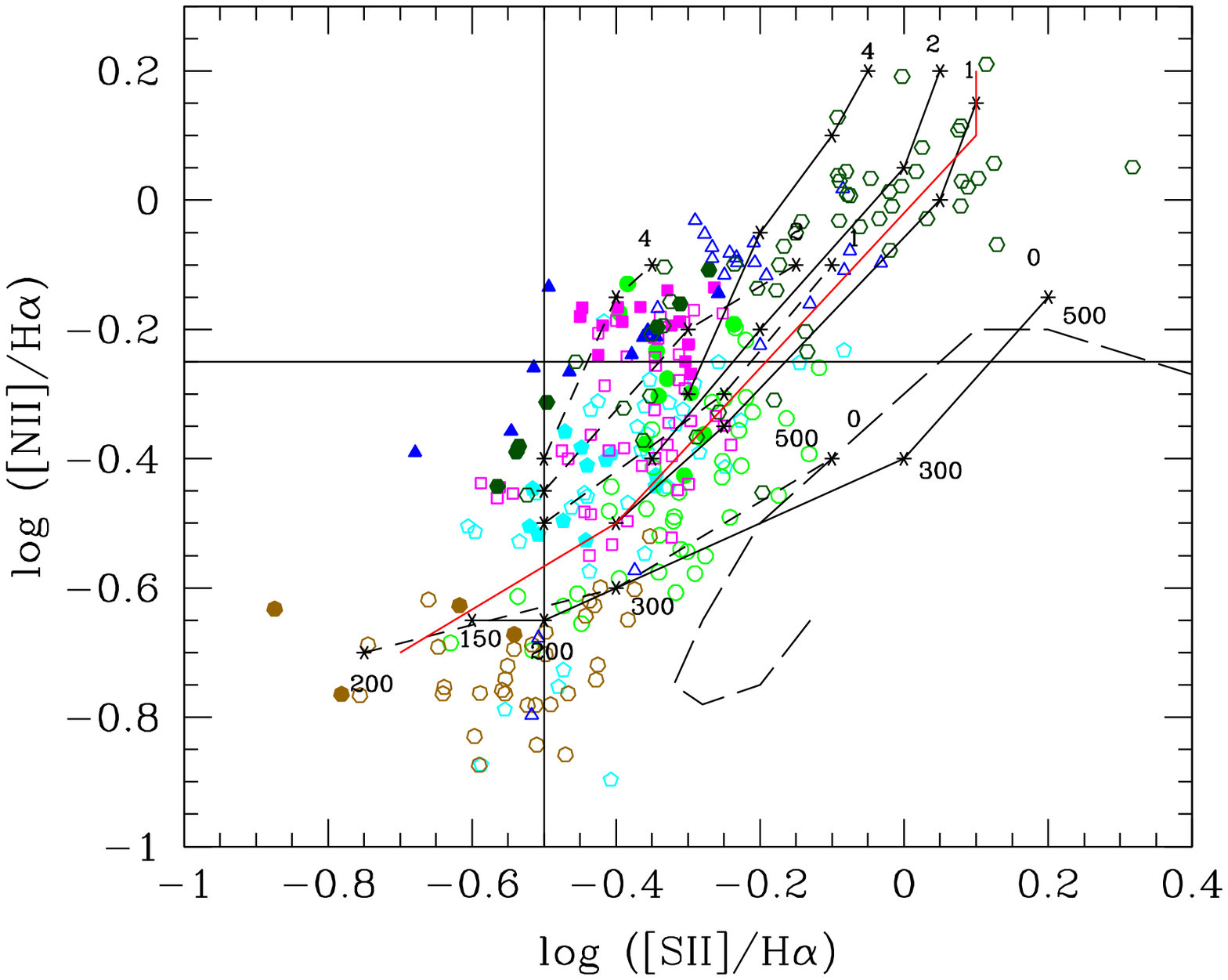} 
\protect\caption[\textsc{[N$\;$ii]}$\lambda$6584/H$\alpha$
vs. \textsc{[S$\;$ii]}$\lambda\lambda$6717,6731/H$\alpha$ 
ratios.]{\textsc{[N$\;$ii]}$\lambda$6584/H$\alpha$
  vs. \textsc{[S$\;$ii]}$\lambda\lambda$6717,6731/H$\alpha$ ratios. This
  graphic can be divided in several regions. In the left bottom corner is
  located the region occupied by typical \textsc{H$\;$ii} regions while top
  right corner of the graphic is the locus for typical LINER-like
  spectrum. Color and shape code is the same as in Figure
  \ref{cociydisp}. Values for the fibers associated to the circumnuclear
  region are indicated with solid symbols while those for the other fibers
  appear with hollow symbols. Models of \citet{dop95} for shocks have been
  superposed. Those without precursor  are indicated with continuous lines
  while those with precursor are plotted using dashed lines. At the beginning
  of each line, it is show the magnetic parameter
  $B/n^{1/2}$~($\mu$G~cm$^{3/2}$). Shocks velocity range from 150 to
  500~km~s$^{-1}$  for models without precursor and from 200 to 500 for models
  with precursor.
  The long-dashed line indicates the \textsc{[N$\;$ii]}$\lambda$6584/H$\alpha$
  and \textsc{[S$\;$ii]}$\lambda\lambda$6717,6731/H$\alpha$ values predicted
  for photoionization by a power-law model for a dusty cloud at $n_e =
  10^3$~cm$^{-3}$ and $Z =Z_\odot$ \citep{gro04}.
   The red
  line indicates the locus for an instantaneous burst model of 4~Myr, $Z =
  Z_\odot$, IMF power-law slope of $-$2.35 and $M_{up} = 100$~M$_\odot$; dust
  effects have been included and $n_e = 10^3$~cm$^{-3}$
  \citep{bar00}. \label{cocicoci}}
\end{figure}

Similarly, the \textsc{[S$\;$ii]}$\lambda\lambda$6717,6731/H$\alpha$ and the
\textsc{[O$\;$i]}$\lambda$6300/H$\alpha$ ratios have also been obtained,
although in smaller field due to poorer signal (the maps are not shown,  
but individual values are presented in Figures 2 and 4). As discussed by
\citet{dop95}, these line ratios, and specially
\textsc{[O$\;$i]}$\lambda$6300/H$\alpha$, 
are more reliable in distinguishing between \textsc{H$\;$ii} and  
LINER like ionization. It is interesting to note that in general these line
ratios indicate an ionization state higher than that inferred from the 
\textsc{[N$\;$ii]}$\lambda$6584/H$\alpha$ ratio. This is shown in Figure
\ref{cocicoci}, which shows the [\textsc{N$\;$ii}]$\lambda$6584/H$\alpha$
vs. [\textsc{S$\;$ii}]$\lambda\lambda$6717,6731/H$\alpha$  
for all the individual spectra/regions of the six systems of the sample
([\textsc{S$\;$ii}]/H$\alpha$ instead of \textsc{[O$\;$i]}/H$\alpha$ has been
selected for this plot since it covers a larger 2D region). In this figure,
vertical and horizontal lines represent the frontier between \textsc{H$\;$ii}
and LINER type of ionization. Many more spectra are classified as LINER
according the [\textsc{S~ii}]$\lambda\lambda$6717,6731/H$\alpha$ ratio
(i.e. points located to the right of the vertical line) than according to
[\textsc{N~ii}]$\lambda$6584/H$\alpha$ (i.e. points above the horizontal
line). In addition, most of the data points are located either in the
LINER-like region according to both line ratios (i.e. upper right quadrant) or
in the region where the [\textsc{S~ii}]$\lambda\lambda$6717,6731/H$\alpha$
ratio is typical of LINER but [\textsc{N~ii}]$\lambda$6584/H$\alpha$ 
typical of \textsc{H$\;$ii} regions (i.e. right bottom quadrant).

For the shake of the following discussion we define circumnuclear regions as
those confined within the central $\sim$  3 arcsec (i.e. r $<$ 1.5 arcsec, and
it is covered by $\sim$ 6 fibers/spectra), and extranuclear regions those which
typically extends for several kpc outwards of this region (i.e. r $>$ 1.5
arcsec). In Fig. 2 we represent the circumnuclear and extranuclear regions
with solid and open symbols, respectively. Note that the circumnuclear region
corresponds roughly with the areas studied previously via long-slit
spectroscopy.  The circumnuclear data in this plot seem to be distributed in a
more compact region, where the extranuclear data cover in general a wider
range of these line quotients. 
  
In order to investigate the different ionization alternatives, the line ratios
predicted by different mechanisms are shown in Figure \ref{cocicoci}. Based in
the apparent 
continuity between LINER and Seyfert spectra, along with the discovery of
X-ray emision and the existence of a wide component in the H$\alpha$ emision
line of some LINERs, photoionization by a power law spectrum coming from an
AGN has been proposed as a possible ionizing mechanism
\citep[e.g.][]{ho93,gro04}. 
Although some of these models could in principle explain
the line ratios measured in the circumnuclear regions, none of the nuclei of
the sample is clearly located in the region identified by these models (see
Figure \ref{cocicoci}, where we have shown one of the models of as example).
In general, the circumnuclear data show either an \textsc{H$\;$ii} like
spectra (indicative 
of an intense star formation) or spectra of compound nature
(LINER+\textsc{H$\;$ii}). This agrees with the classification in the
mid-infrared for these objects \citep{tan99,rig99} that includes them in the
group of \emph{starburst} using the line-to-continuum ratio of PAH at
$7.7$~$\mu$m.

Regarding the extranuclear regions, the AGN models \citep{ho93,gro04} are not,
in 
general,  likely to be representative of these low-density ($n_e < 10^{3}
$~cm$^{-3}$) regions, as it is also indicated by the relatively small fraction
of data points located within the area defined by these models. However it is
interesting to note that the case of IRAS~17206$-$0014 may represent an
exception in this context (see discussion in 4.2). In short, although we
cannot discard a possible contribution of AGN-like ionization in some regions,
clearly this mechanism cannot explain in general the observed lines ratios
represented in Figure 2.  

Ionization by young stars could be an obvious alternative mechanism to explain
the line ratios. \citet{bar00} have shown
that starburst models during the Wolf-Rayet dominated phase can explain the
spectra of some LINERs, but only under very restricted conditions. In Figure  
\ref{cocicoci} we have plotted the Barth \& Shields' model that best fits the
locus of our data as a red line. This corresponds to an instantaneous burst
model of 4~Myr, $Z = Z_\odot$, Initial Mass Function (IMF) power-law  
slope of $-$2.35 and $M_{up} = 100$~M$_\odot$ and an interstellar medium
characterized by an electron density ($n_e$) of 10$^3$~cm$^{-3}$. These
conditions are very unlikely to be representative of the extranuclear regions
of all these ULIRGs, especially taking into account the relatively young and
short-lived population involved (i.e. for clusters younger than 3~Myr or older
than 6 Myr, and for models with a constant star formation rate, the softer
ionizing continuum results in an emission spectrum more typical of
\textsc{H$\;$ii} regions. Furthermore, the fraction of ULIRGs with hints of WR
signatures in their spectrum (i.e. broad optical feature at 4660 \AA) is less
than 10\% \citep{arm89}.  

The most likely mechanism to explain the observed ionization in the extended,
extranuclear regions is the presence of large scale shocks. Figure 2 presents
the predicted line ratios for a representative set of shock models
\citep{dop95}. These ratios agree with the range of observed values for shock
velocities of 150 km s$^{-1}$ to 500 km s$^{-1}$ in either a neutral
(continuous lines), or a pre-ionized medium (dashed lines). Moreover, such
high speed flows are routinely detected in the extranuclear regions of ULIRGs
as shown by detailed two-dimensional kinematic studies \citep{col05}. Velocity
fields inconsistent in general with ordered motions, and 
with typical peak-to-peak velocities of 200 to 400  km s$^{-1}$ are detected
in the tidal tails, and extranuclear regions of ULIRGs on scales of few to
several kpc away from the nucleus, and almost independent of the dynamical
phase of the merger \citep[see][and references
therein]{col04,col05}. Moreover, the presence of highly turbulent gas, as
identified by large velocity dispersions of 70 to 200 km s$^{-1}$ (Colina et
al. 2005), further supports the scenario of fast shocks as the main ionization
mechanism in these regions.  

In summary, the ionization of the extranuclear regions in the ULIRGs studied
here, can be hardly explained by accretion powered AGN or by young starbursts,
but is consistent with fast, large scale shocks.

\subsection{Excitation and Velocity Dispersions: Further Evidence for
  Ionization by Shocks }  

The positive correlation between the velocity dispersion and ionization found
by some authors using circumnuclear (slit) spectra of ULIRGs has been
considered as further evidence supporting the presence of shocks in these
objects \citep{arm89,dop95,vei95}. The present study also supports the
correlation previously observed. As shown in Figure 3, the
\textsc{[S$\;$ii]}$\lambda\lambda$6717,6731/H$\alpha$ and velocity dispersion
values derived from the integrated spectra -- i.e. combining the individual
spectra for each object, are consistent with previous results
\citep{arm89}. However, these spectra are not necessarily representative of
the extranuclear regions. 

In  Figure \ref{cociydisp} we present similar plots, but now each data point
represents the value for a specific spectrum (fiber) (i.e. different position
in the extranuclear nebula) but excluding the circumnuclear region, for each
individual galaxy, and using three different line ratios 
(\textsc{[N$\;$ii]}$\lambda$6584/H$\alpha$,
\textsc{[S$\;$ii]}$\lambda\lambda$6717,6731/H$\alpha$, 
\textsc{[O$\;$i]}$\lambda$6300/H$\alpha$). The dashed horizontal lines
indicate the borderline between \textsc{H$\;$ii} and LINER ionization. In the
top panels the data for the individual galaxies (except IRAS 17208$-$0014, see
below) are combined. These panels indicate that, while the correlation of the
velocity dispersions with the line-ratio is not so well defined for
\textsc{[N$\;$ii]}$\lambda$6584/H$\alpha$, 
for the other two line ratios, and specially
\textsc{[O$\;$i]}$\lambda$6300/H$\alpha$ 
(i.e. the most reliable diagnostic ratio to detect ionization by shocks
according to models by Dopita \& Sutherland, 1995), the correlation is
clear. The fact that the extranuclear data of these five systems follow a
well defined relation between the line-ratio and the velocity dispersion
reinforces the idea that shocks are also the dominant ionization source at
large scales ($>$ 2-3 kpc).
Individually, two systems, IRAS~12112+0305 and IRAS~14348$-$1447, show a clear
correlation in the three line ratios. For three of the remaining systems,
IRAS~08572+3915,IRAS~15206+3342 and IRAS~15250+3609, the range in velocity
dispersion is relatively small to show the correlation.

Finally, IRAS~17208$-$0014 does not follow the mean behavior observed in the
other systems, showing a wider range of line 
ratio values.  This galaxy has been studied in detail by \citet{arr03} and
\citet{col05}, and it is the only clear case in this sample
showing rotation on scales of several kpc. This may be an indication that this
system is in a different (probably more evolved) dynamical phase and/or it has
had a different merging history.  At any event, the fact that the gas
kinematics indicates a more relaxed and virialized system suggests that shocks
are not the dominant ionization mechanism in this galaxy 
and, therefore the above mentioned correlation should not be expected. For
this galaxy the origin of the LINER-like ionization in the extranuclear region
should be different (note that the three line ratios shown in Fig.~4 are
consistent with LINER-like ionization). A hint that this is the case comes
from the detection of an extended ($\sim$ 4 kpc)  hard X-ray nebula in this  
galaxy \citep{pta03}, which would provide an ionizing spectrum similar to 
that of an AGN. This may explain the fact that the excitation  
of this object is higher than that of the rest of the galaxies and similar to
that expected from a low luminosity AGN \citep[Fig.2;][]{ho93}.   

\subsection{Origin of the Shocks in the Circumnuclear and Extranuclear
  Regions: Superwinds and Merger Induced Flows}

In previous sections shocks have been identified as the main ionization
mechanism  in the extended, extranuclear ionized regions. Moreover, the
detection of a positive correlation between the ionization status of  the gas,
as best indicated by the shock tracer \textsc{[O$\;$i]}$\lambda$6300/H$\alpha$
ratio, and the velocity  dispersion of the gas, suggests a direct causal
relation between the LINER ionization and the presence of shocks. What is the
origin of the shocks in the circumnuclear, and in the extranuclear regions
extended at distances of up to 10-15 kpc from the nucleus? Some authors have
found evidence supporting the existence of so-called superwinds generated by
the combined effect of massive stars winds and supernova explosions in intense
nuclear starbursts \citep{hec90}. These superwinds, identified by the presence
of kinematically distinct components in the profiles of the emission
\citep{hec90} or absorption lines \citep{mar05,rup05a,rup05b}, generate shocks
in the circumnuclear regions as the stellar winds move through the
interstellar medium.  

Recent studies in a large sample of LIRGs and ULIRGs conclude that the
presence of superwinds has to be an almost universal phenomenon in the
circumnuclear regions of ULIRGs (typical angular sizes of about of
1$^{\prime\prime}$ or 
1 to 2~kpc, depending on redshift) as kinematically distinct components of the
neutral interstellar NaD lines, blueshifted by a median velocity of 350 km
s$^{-1}$ are detected in at least 70\% of ULIRGs
\citep{mar05,rup05a,rup05b}. These velocity components are also detected in
our integral field 
spectra for some systems. Out of the six galaxies in the sample, our data show
the presence of double H$\alpha$ line profiles in the circumnuclear regions of
the northern and southern nucleus of the interacting pairs IRAS 12112+0305 and
IRAS 14347$-$1448 (see Figure 5). These secondary velocity components are
blueshifted with respect to system by 150 km s$^{-1}$ and 300 km s$^{-1}$,
respectively. In addition to these galaxies, similar signatures have been
identified in the circumnuclear regions of more evolved ULIRGs such as IRAS
15250+3609 ($V-V_{sys} =-170$~km~s$^{-1}$, Monreal-Ibero 2004), and Arp~220
\citep[peak-to-peak velocity of 1000 km s$^{-1}$,][]{arr01}. However, our
IFS data show that the presence of double components, when detected, is always
confined to the nuclear and circumnuclear regions, i.e. distances of 1 to 2
kpc from the nucleus. The lack of detection of double components in the
extranuclear regions, at distances of several kpc away from the nucleus, can
be interpreted as if the high velocity outflows associated with the nuclear
superwinds were not present at these distances, or as if they were of much
lower amplitude (less than 100 km s$^{-1}$), therefore not been detected as
kinematically distinct components with the present spectral resolution.  

On the other hand, the complex
two-dimensional velocity field and velocity dispersion structure of the
extranuclear ionized regions of ULIRGs \citep{col04,col05} shows in general
large velocity gradients with peak-to-peak velocities of few to several
hundreds km s$^{-1}$ associated with tidal tails and extranuclear regions at
distances of several kpc away from the massive circumnuclear
starbursts. Moreover, the largest values of the velocity dispersion in many 
ULIRGs (up to 200 km s$^{-1}$) are detected not in the nucleus but in
extranuclear regions \citep{col05}, implying therefore the presence of an
extended, highly turbulent medium on kpc-size scales. As shown by specific
models of the nearest ULIRG, Arp 220, tidally induced flows lead to relative
gas velocities that are much larger than the original impact velocities of the
galaxies \citep{mcd03}, and therefore high speed flows of several hundreds km
s$^{-1}$ are a natural consequence of the merging process associated with
ULIRGs. The presence of these tidally induced, high velocity flows and highly
turbulent medium will generate shocks that in turn will heat and ionize the
interstellar medium producing the observed LINER type spectra as in the
nearest ULIRG, Arp 220 \citep{mcd03,col04}. 

In short, the lack of superwind signatures and the kinematic properties of the
gas in the extranuclear regions, supports the idea that merger induced flows
are the origin of the fast shocks producing the LINER-like excitation in these
extended regions. 

\section{CONCLUSIONS}

Integral Field Spectroscopy with the INTEGRAL fiber system has been used to
analyze the circumnuclear and extranuclear ionization structure of six
low-$z$ ULIRGs. The main results can be summarized as follows: 

\begin{enumerate}

\item The two-dimensional ionization characteristics of the extranuclear
  regions of these galaxies correspond to these of LINERs. This is clearly
  indicated by the \textsc{[S$\;$ii]}$\lambda\lambda$6717,6731/H$\alpha$ and
  especially the \textsc{[O$\;$i]}$\lambda$6300/H$\alpha$ line ratios which
  allow 
  to discriminate reliably between the \textsc{H$\;$ii} and LINER ionization
  in low excitation conditions (i.e. \textsc{[O$\;$iii]}$\lambda$5007/H$\beta$
  $\leq$ 2.5).   

\item The observed LINER-type line ratios in the extranuclear regions are in
  general better explained with ionization by fast shocks with velocities of
  150 to 500 km s$^{-1}$, rather than with AGN or starburst
  photoionization. Further evidence  pointing to shocks as the dominant source
  of ionization comes from a positive correlation between the  ionization
  state and the  velocity dispersion of the ionized gas. The present
  two-dimensional data  show that this correlation  holds especially if the
  \textsc{[O$\;$i]}$\lambda$6300/H$\alpha$ line ratio is used.   

\item Although signatures for superwinds are observed in the circumnuclear
  regions of some systems,  no kinematic evidence for such a mechanism is
  found in the extranuclear regions. Alternatively, the shocks that produce
  the observed LINER-type ionization in the extranuclear regions could be due
  to a different phenomenon. Taking into account the general 2D kinematic
  characteristics of the extranuclear regions in these objects, which indicate
  disordered motions with peak-to-peak velocities of about 400 km~s$^{-1}$, and
  velocity dispersions of up to 200 km s$^{-1}$, the origin of the shocks are
  most likely caused by tidally induced large scale flows produced during the
  merging process. 

\item The galaxy IRAS 17208$-$0014 presents a peculiar kinematical and
  ionization structure.  For this galaxy the origin of the LINER-type
  ionization in the extranuclear region is most likely explained by the
  presence of the recently detected hard X-ray extended (4 kpc) emission that
  would produce an ionizing spectrum similar to that of an AGN. This may
  explain the fact that the excitation of this object is higher than that of
  the rest of the galaxies, and compatible with that expected in low
  luminosity AGNs.    

\end{enumerate}

\acknowledgments

AMI acknowledges support from the Euro3D Research Training Network,
funded by the EC (HPRN-CT-2002-00305). Financial support was provided
by the Spanish Ministry for Education and Science through grant
AYA2002-01055. Work based on observations with the William Herschel Telescope
operated on the island of La Palma by the ING in the Spanish Observatorio del
Roque de los Muchachos of the Instituto de Astrof\'{\i}sica de Canarias.

\clearpage

\clearpage

\clearpage

\begin{figure}[htb!]
\centering
\epsscale{1.00}
\includegraphics[width=13cm,angle=0, clip=,bbllx=35, bblly=210, bburx=560,
bbury=645]{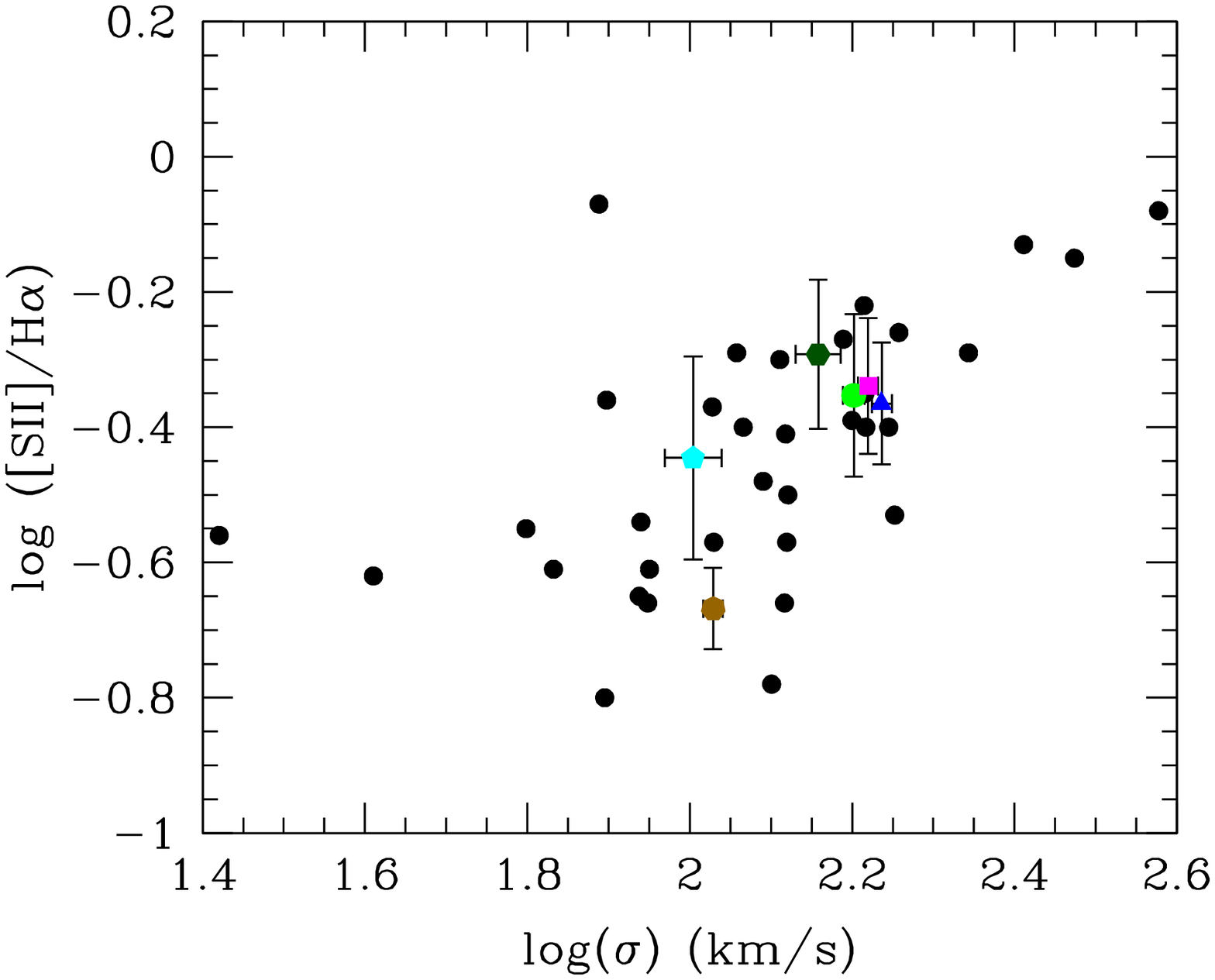} 

\protect\caption{Line width-line intensity correlation for LIRGs. Our data are
  plotted in color while black points are taken from \citet{arm89}. The
  color/symbol code used in all figures in this paper is following:
  IRAS~08572+3915, cyan pentagon; IRAS~12112+0305, 
  green circle; IRAS~14348$-$1447 magenta square; IRAS~15206+3342, brown
  heptagon; IRAS~15250+3609, blue triangle; IRAS~17208$-$0014, olive
  hexagon. Error bars were derived from the errors in the line
  fitting. \label{fig_total}}  
\end{figure}

\clearpage

\begin{figure}[htb!]
\centering
\begin{tabular}{ccc}
\includegraphics[width=3.6cm,angle=0, clip=,bbllx=30, bblly=275, bburx=560,
bbury=640]{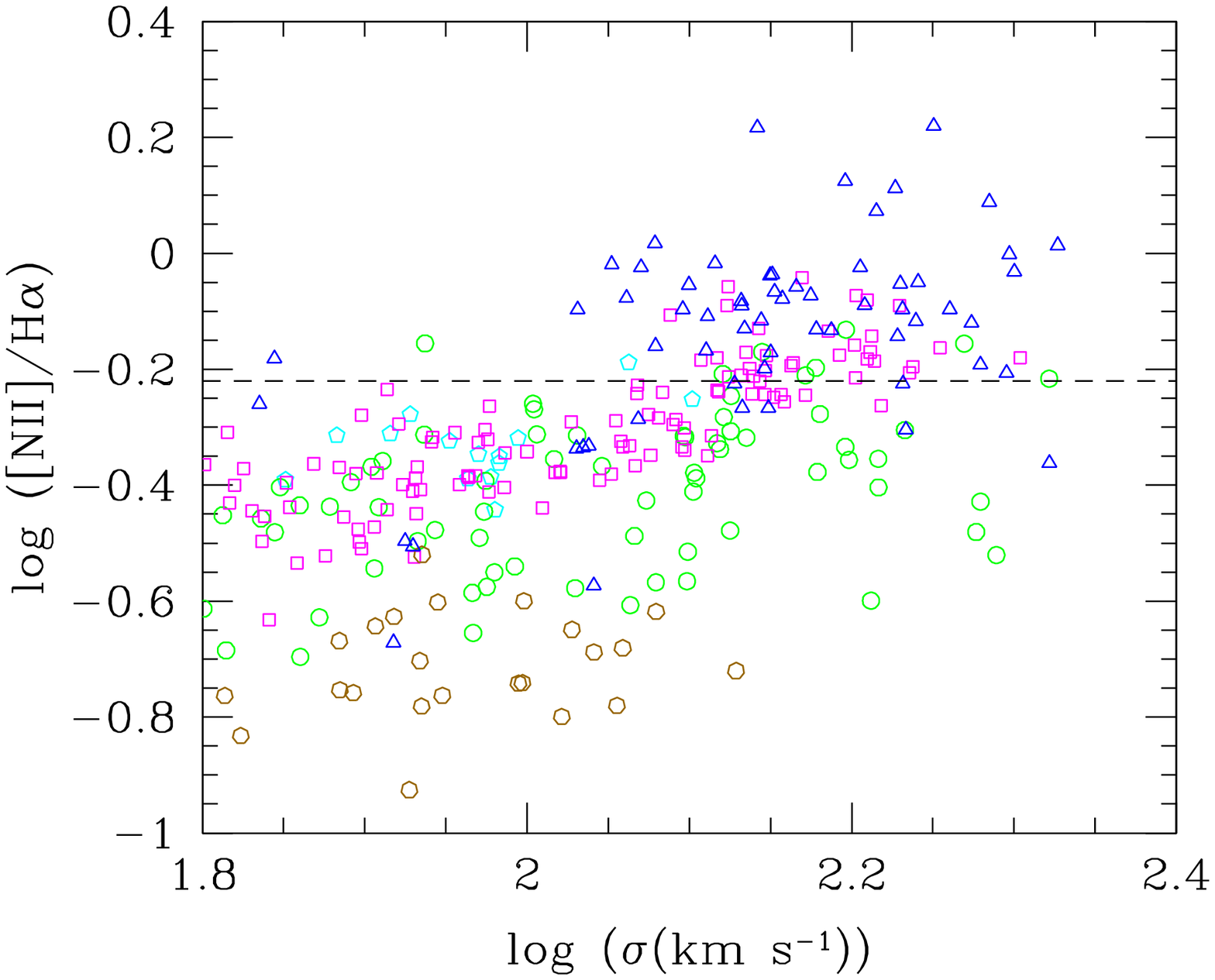} &
\includegraphics[width=3.6cm,angle=0, clip=,bbllx=30, bblly=275, bburx=560,
bbury=640]{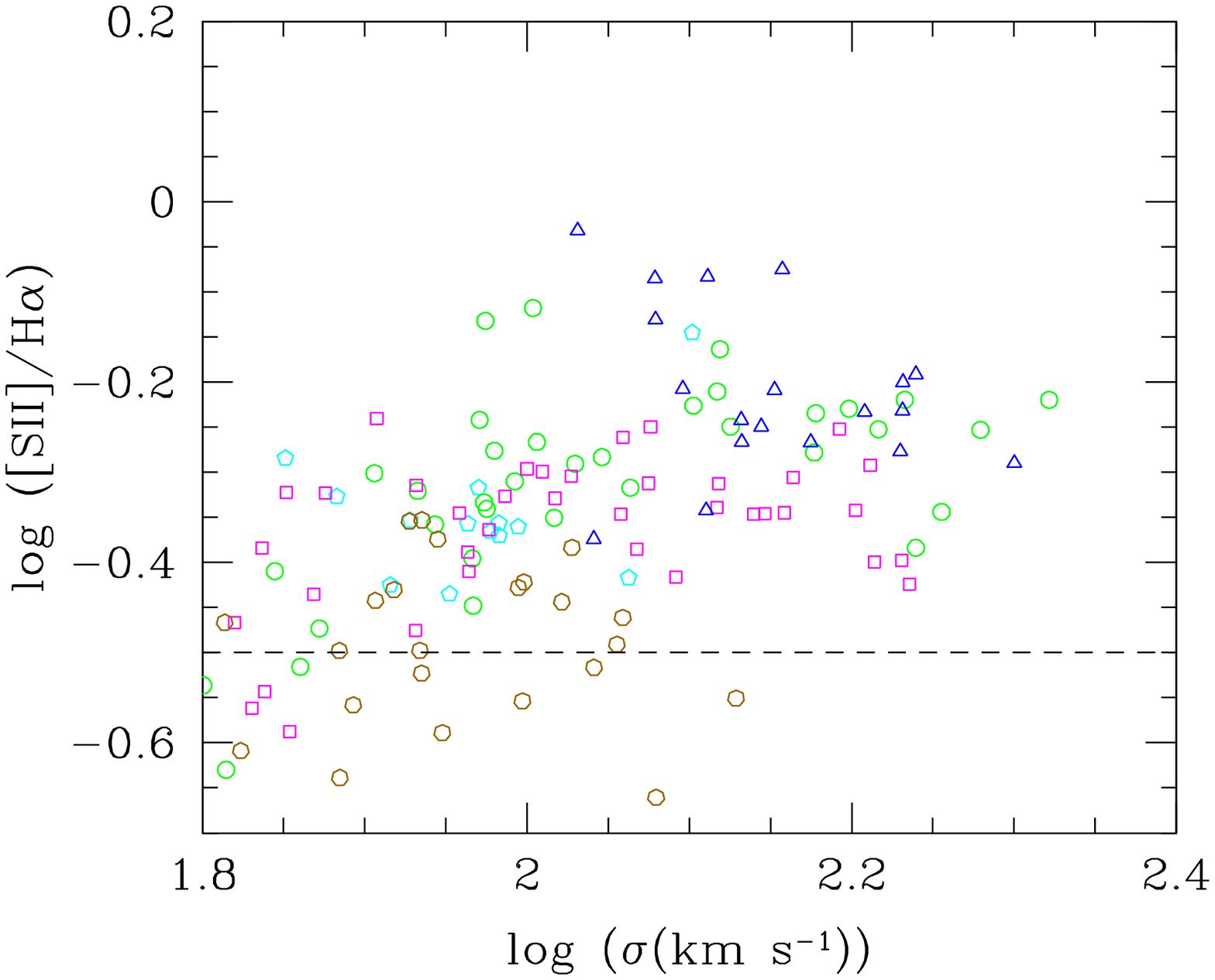} &
\includegraphics[width=3.6cm,angle=0, clip=,bbllx=30, bblly=275, bburx=560,
bbury=640]{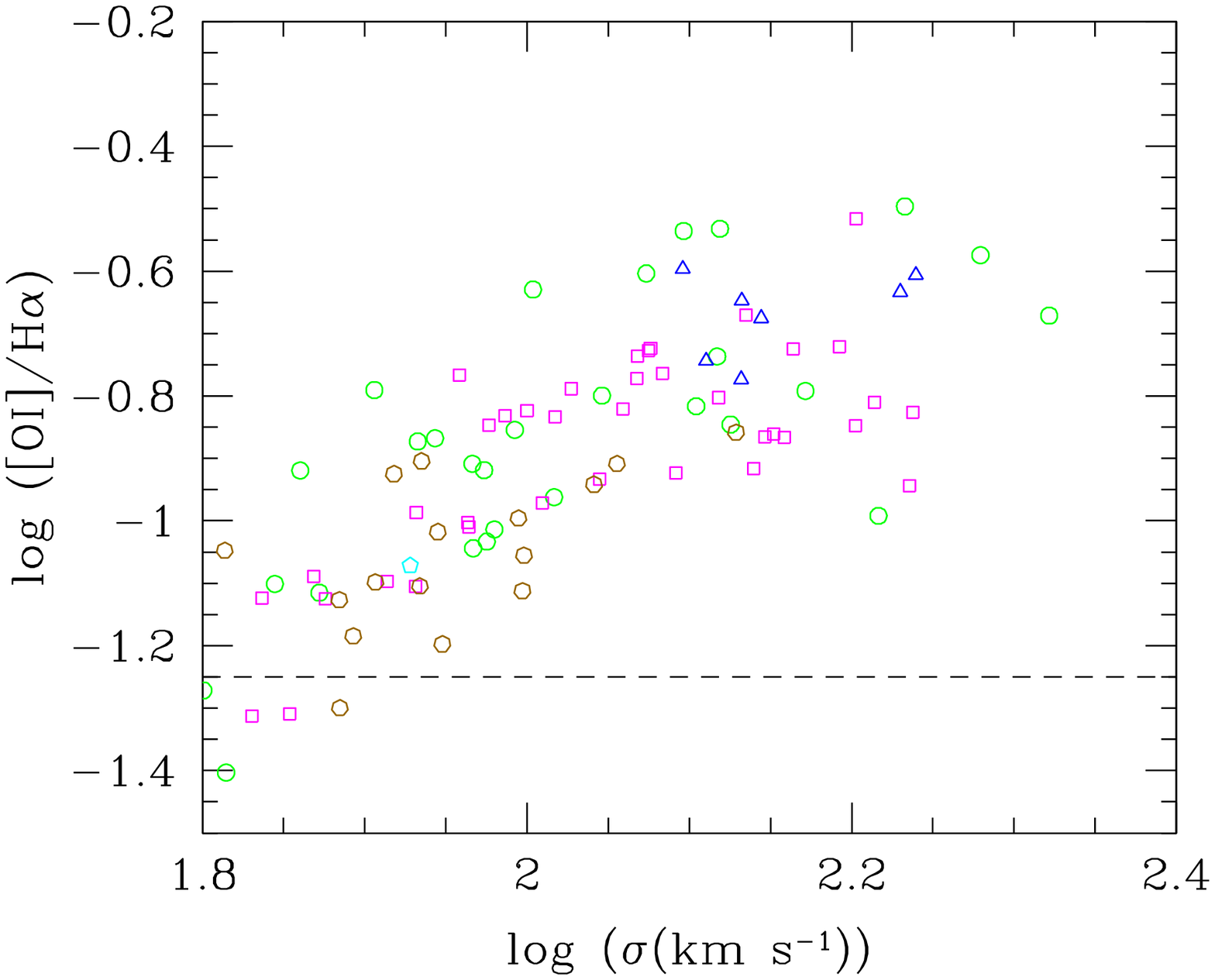}\\ 
\includegraphics[width=3.6cm,angle=0, clip=,bbllx=30, bblly=275, bburx=560,
bbury=640]{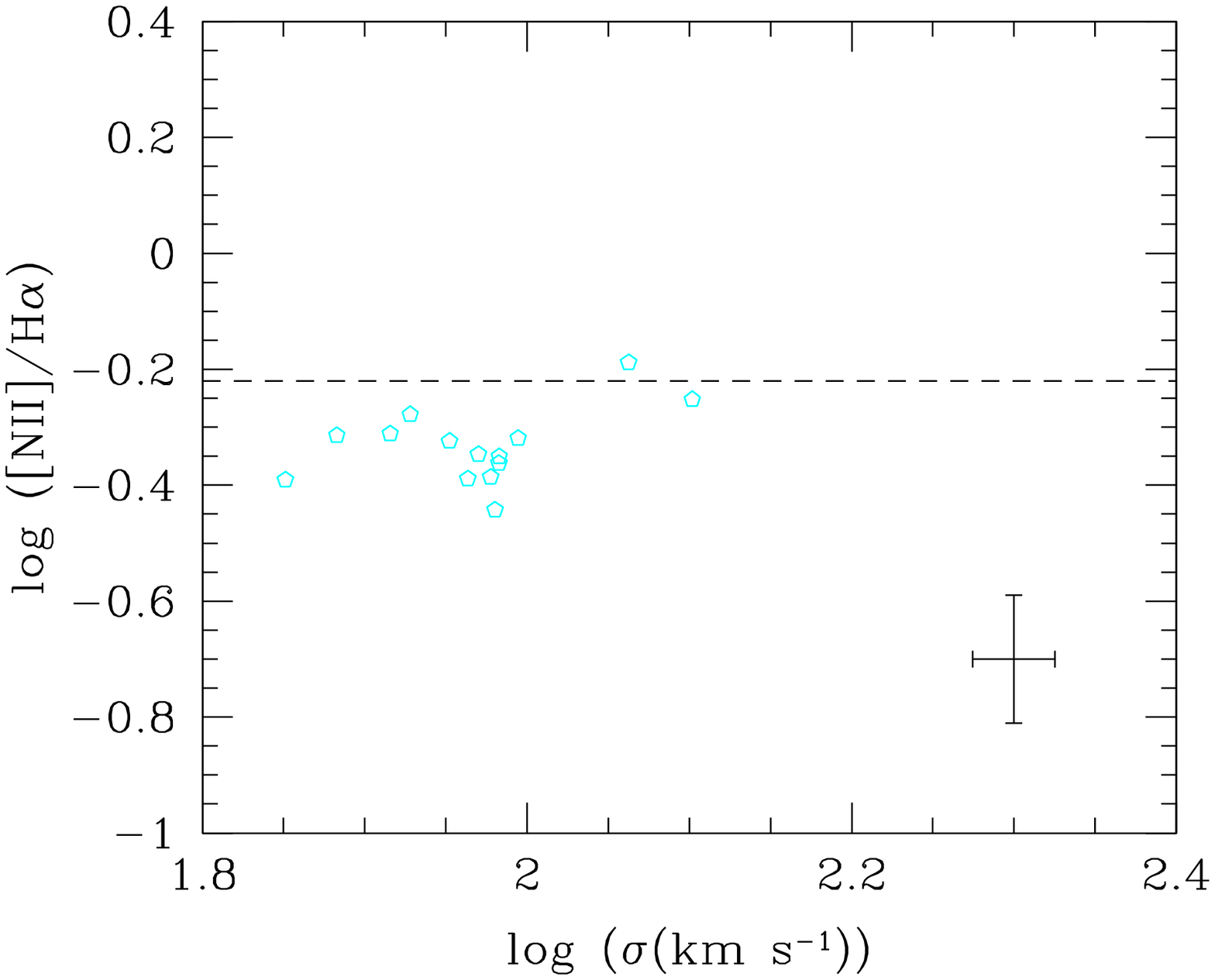} &
\includegraphics[width=3.6cm,angle=0, clip=,bbllx=30, bblly=275, bburx=560,
bbury=640]{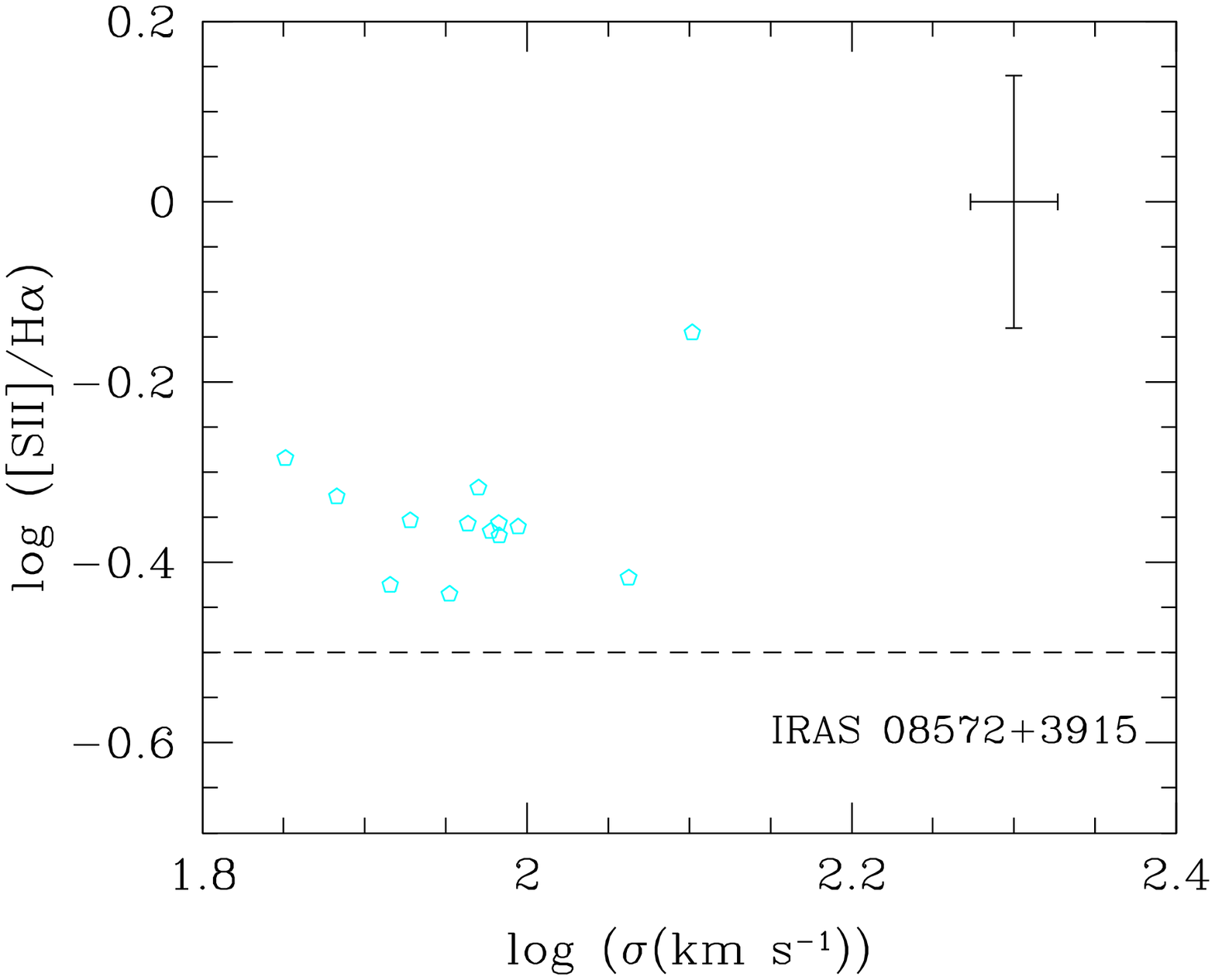} &
\includegraphics[width=3.6cm,angle=0, clip=,bbllx=30, bblly=275, bburx=560,
bbury=640]{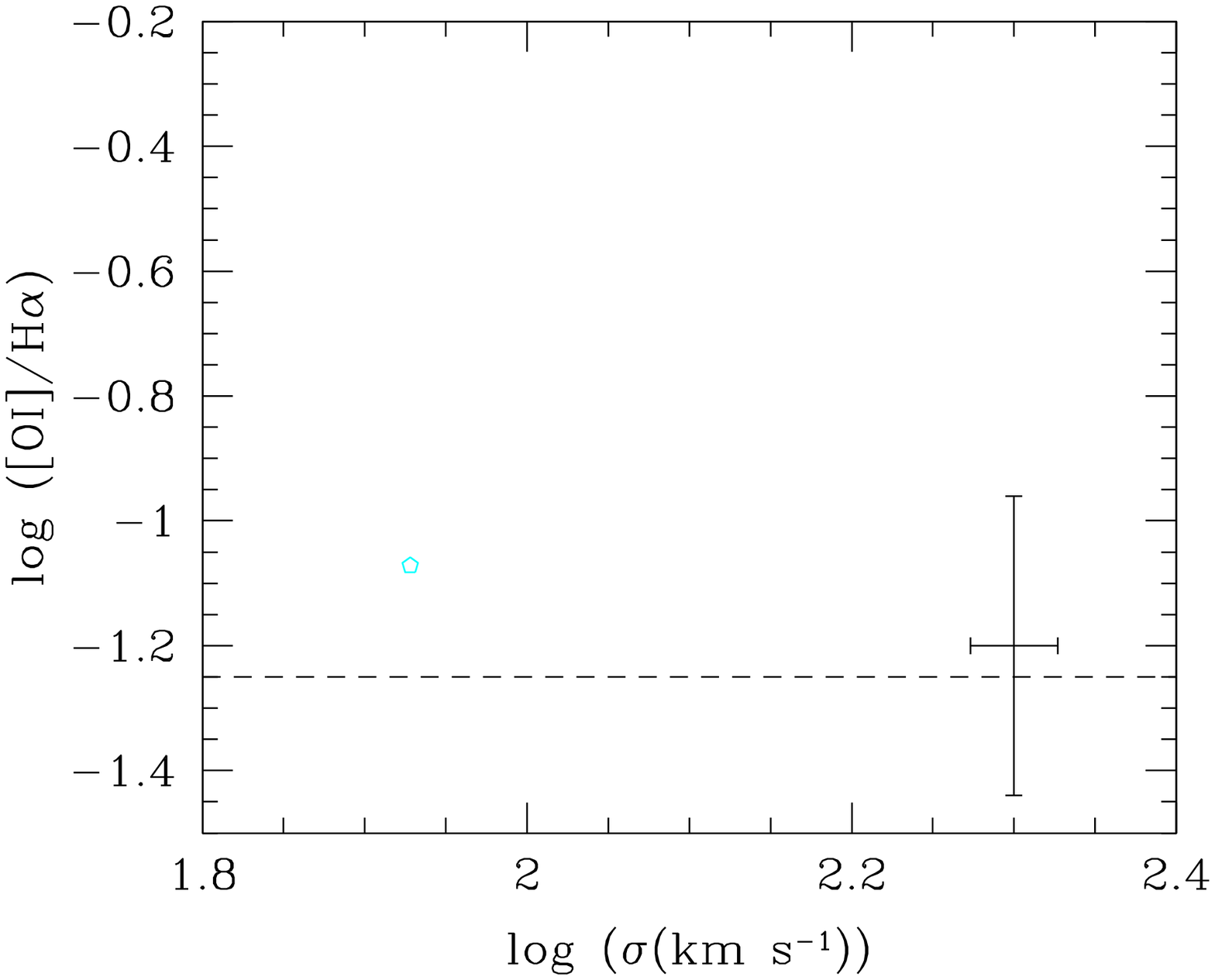}\\ 
\includegraphics[width=3.6cm,angle=0, clip=,bbllx=30, bblly=275, bburx=560,
bbury=640]{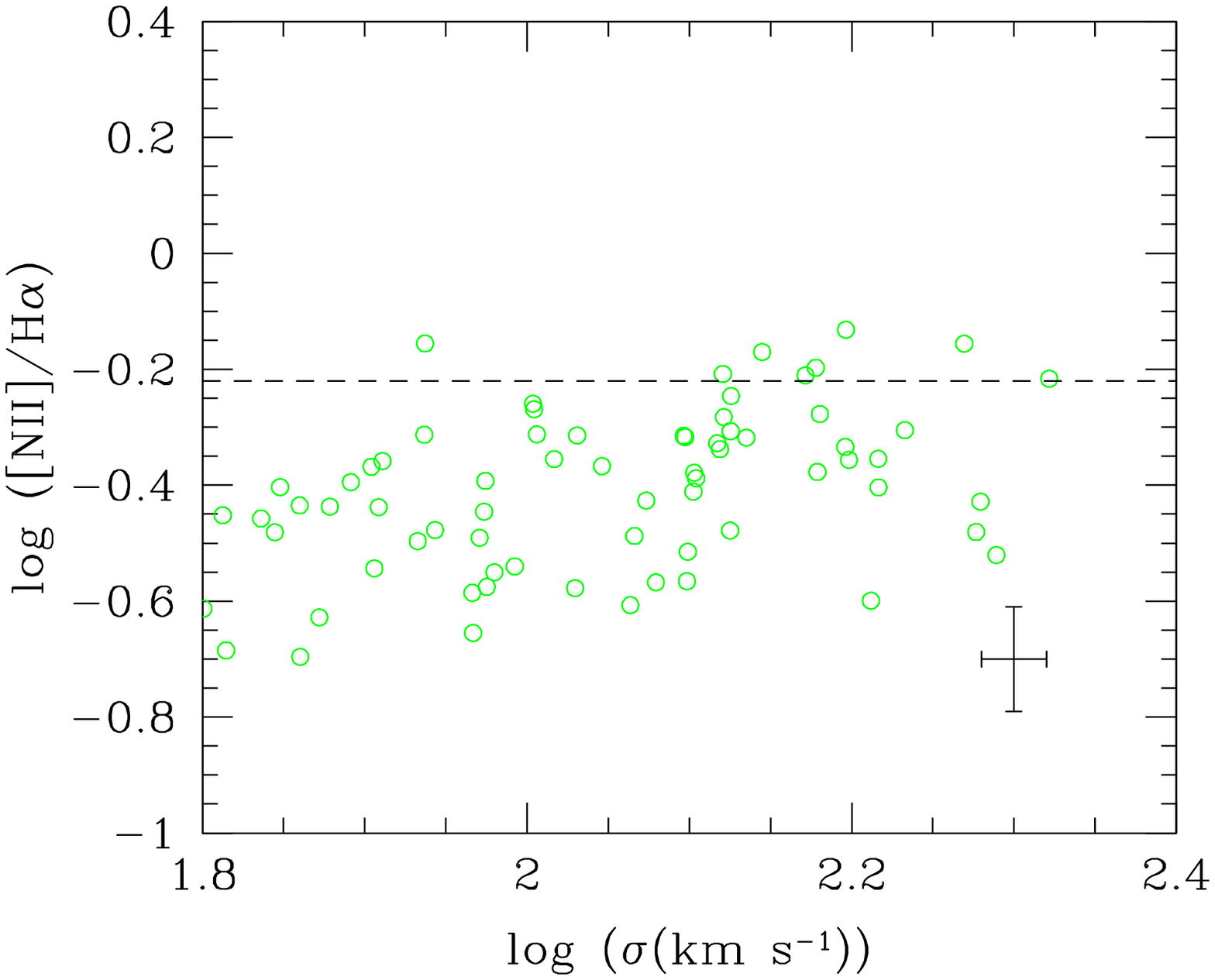} &
\includegraphics[width=3.6cm,angle=0, clip=,bbllx=30, bblly=275, bburx=560,
bbury=640]{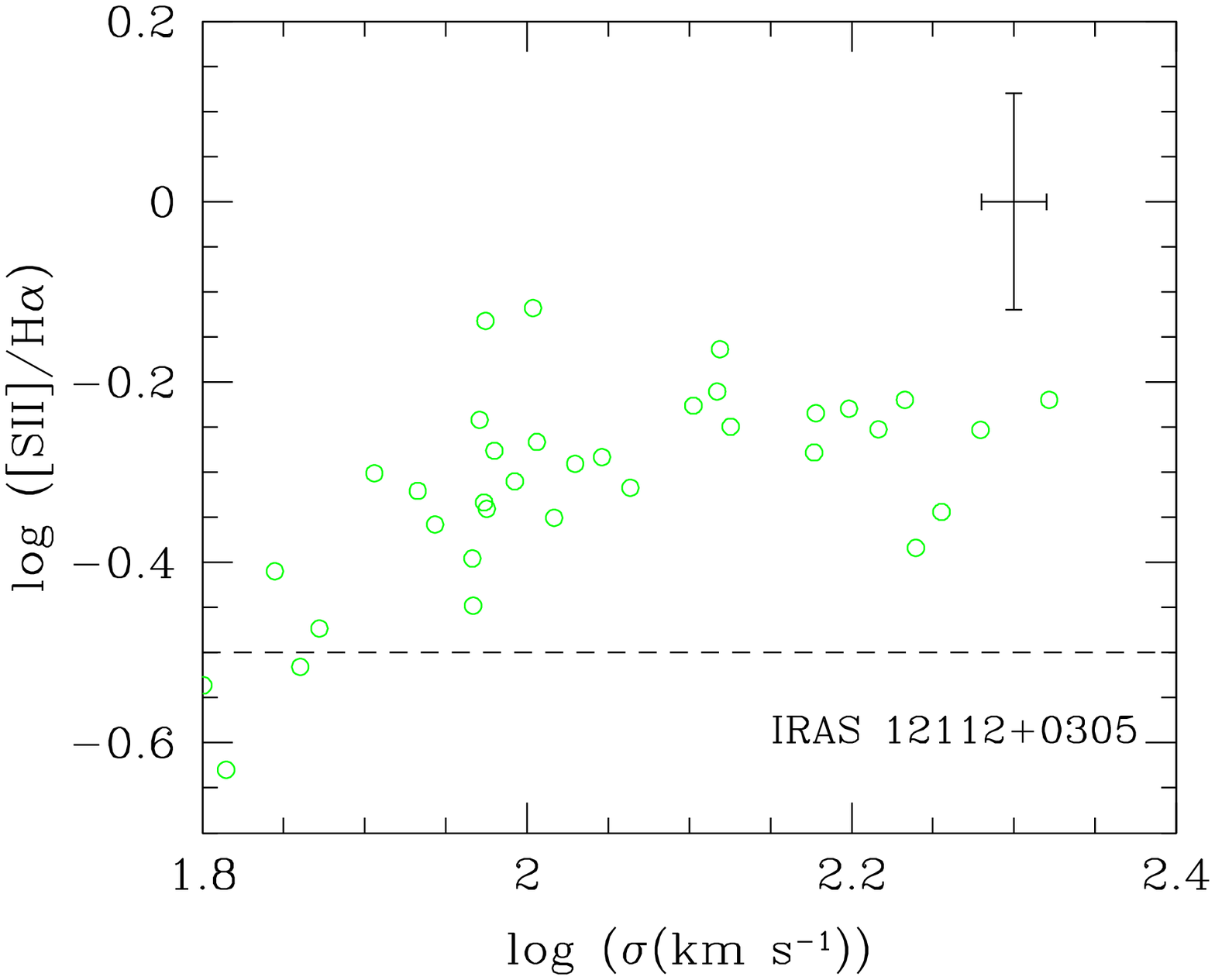} &
\includegraphics[width=3.6cm,angle=0, clip=,bbllx=30, bblly=275, bburx=560,
bbury=640]{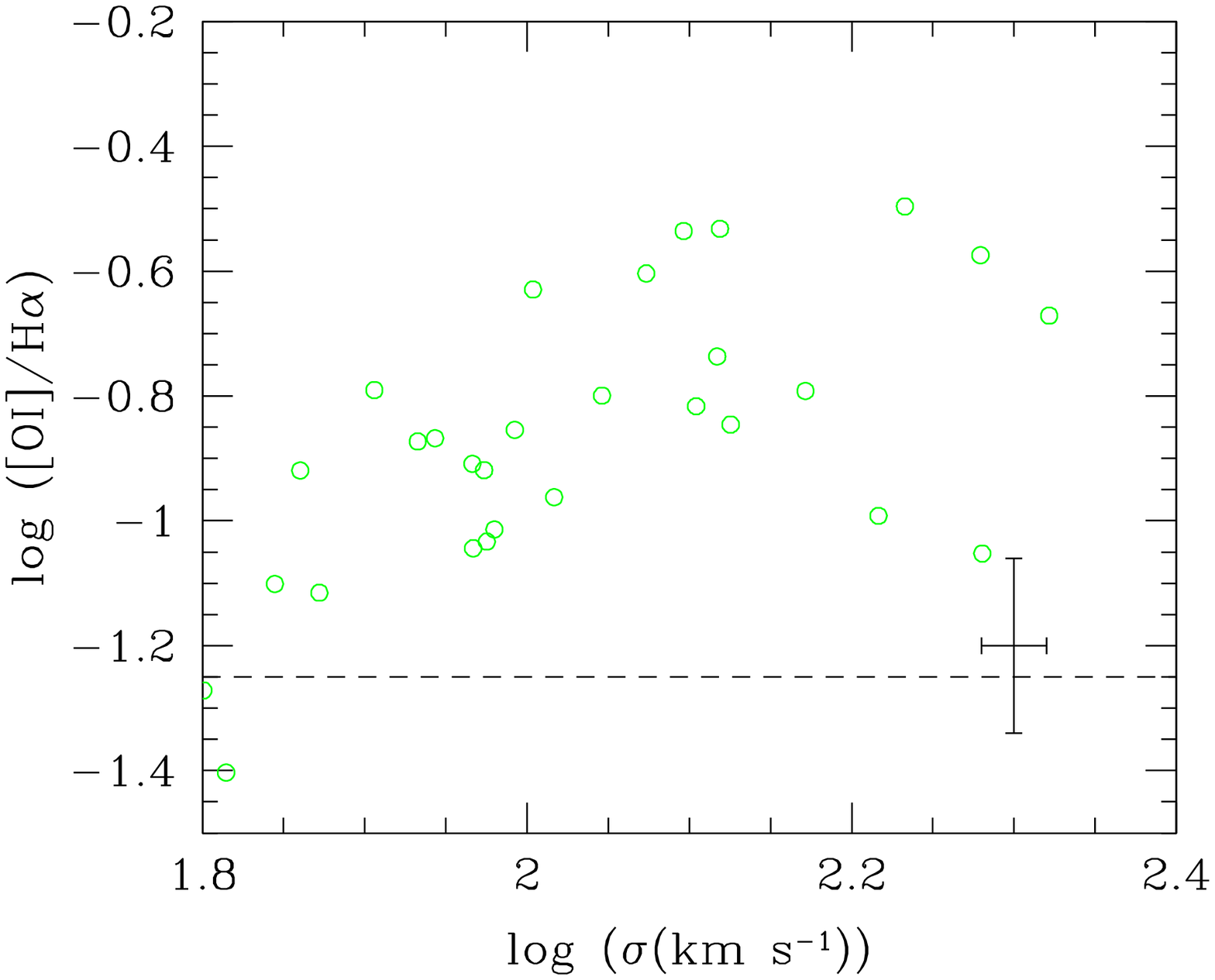}\\
\includegraphics[width=3.6cm,angle=0, clip=,bbllx=30, bblly=275, bburx=560,
bbury=640]{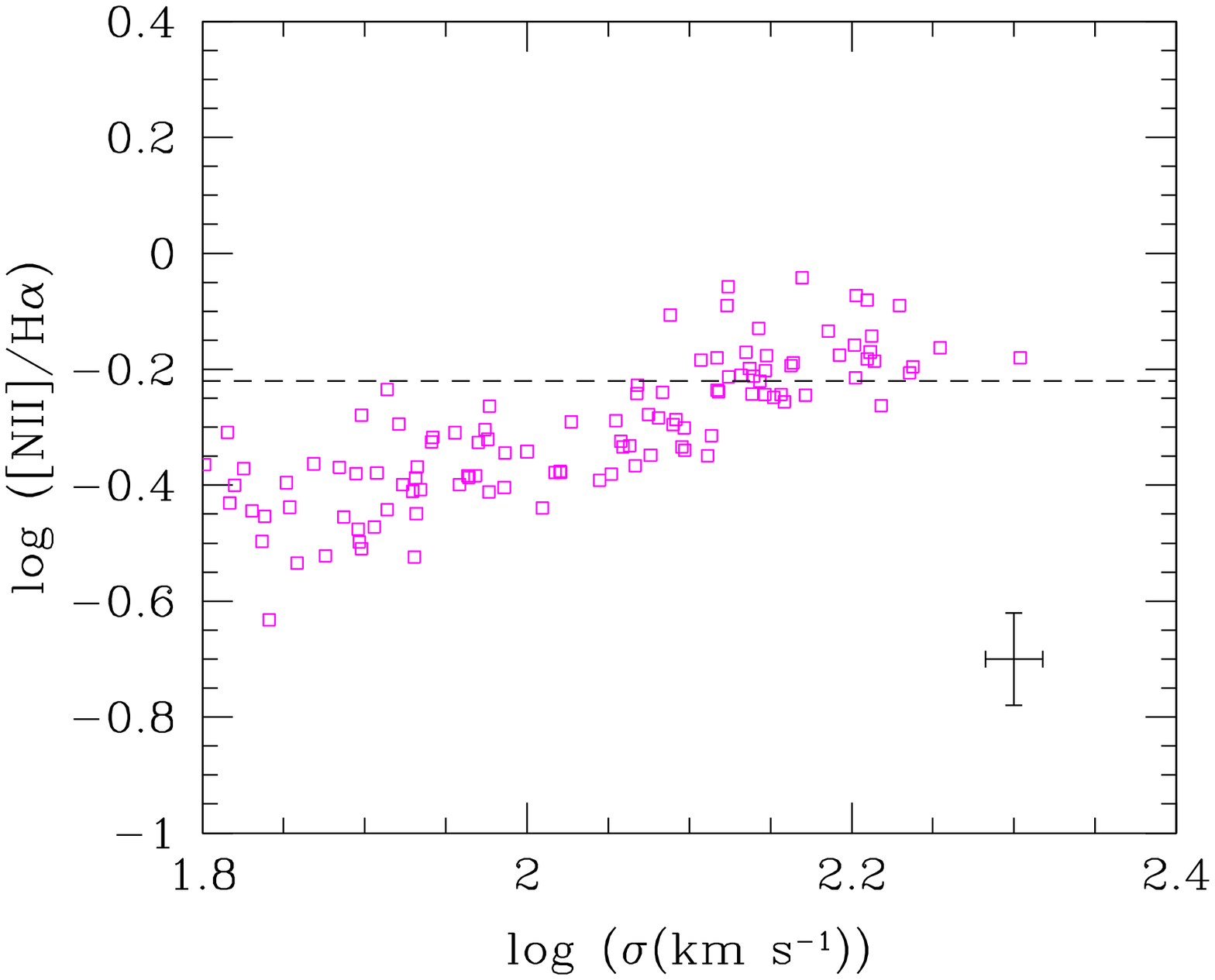} &
\includegraphics[width=3.6cm,angle=0, clip=,bbllx=30, bblly=275, bburx=560,
bbury=640]{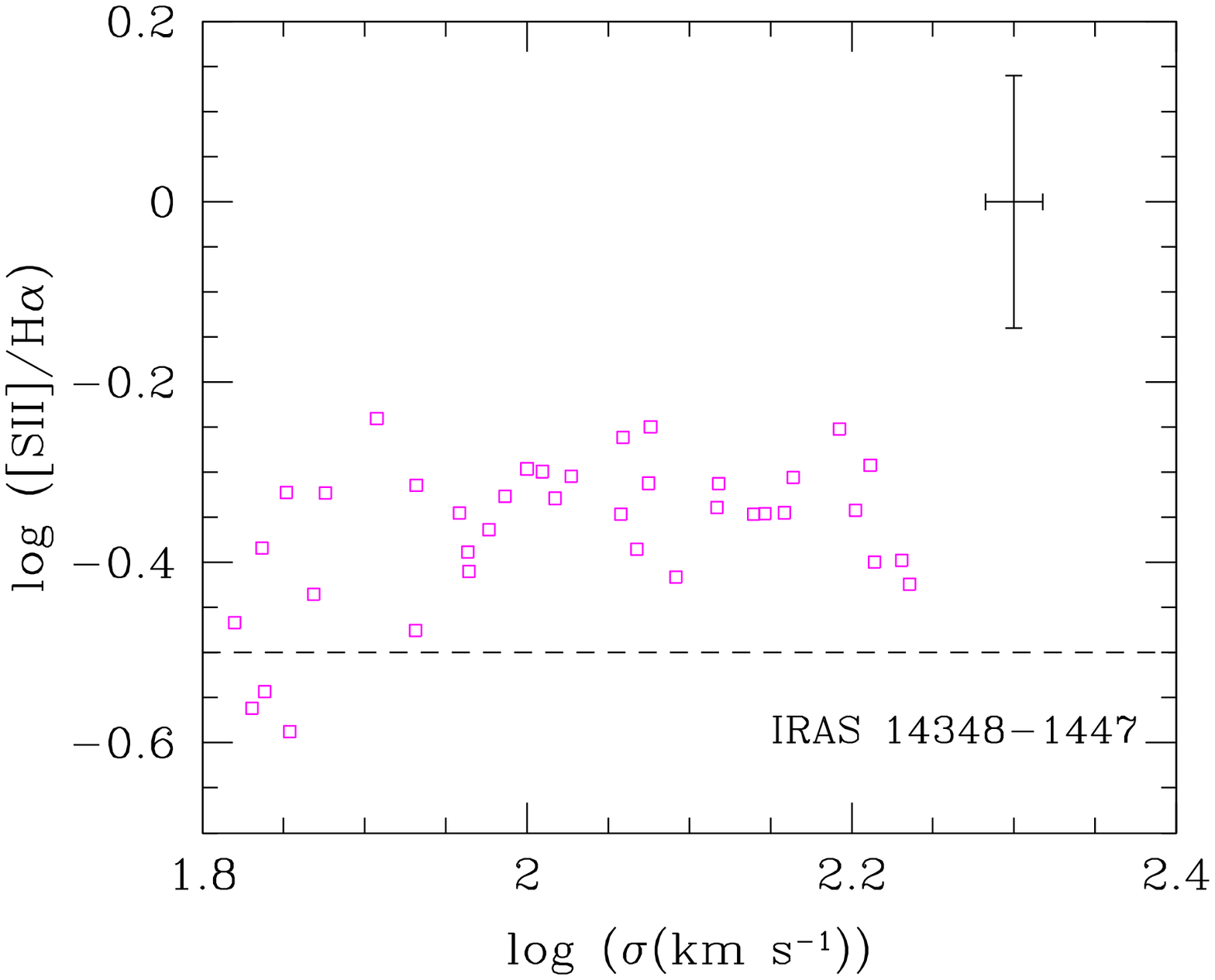} &
\includegraphics[width=3.6cm,angle=0, clip=,bbllx=30, bblly=275, bburx=560,
bbury=640]{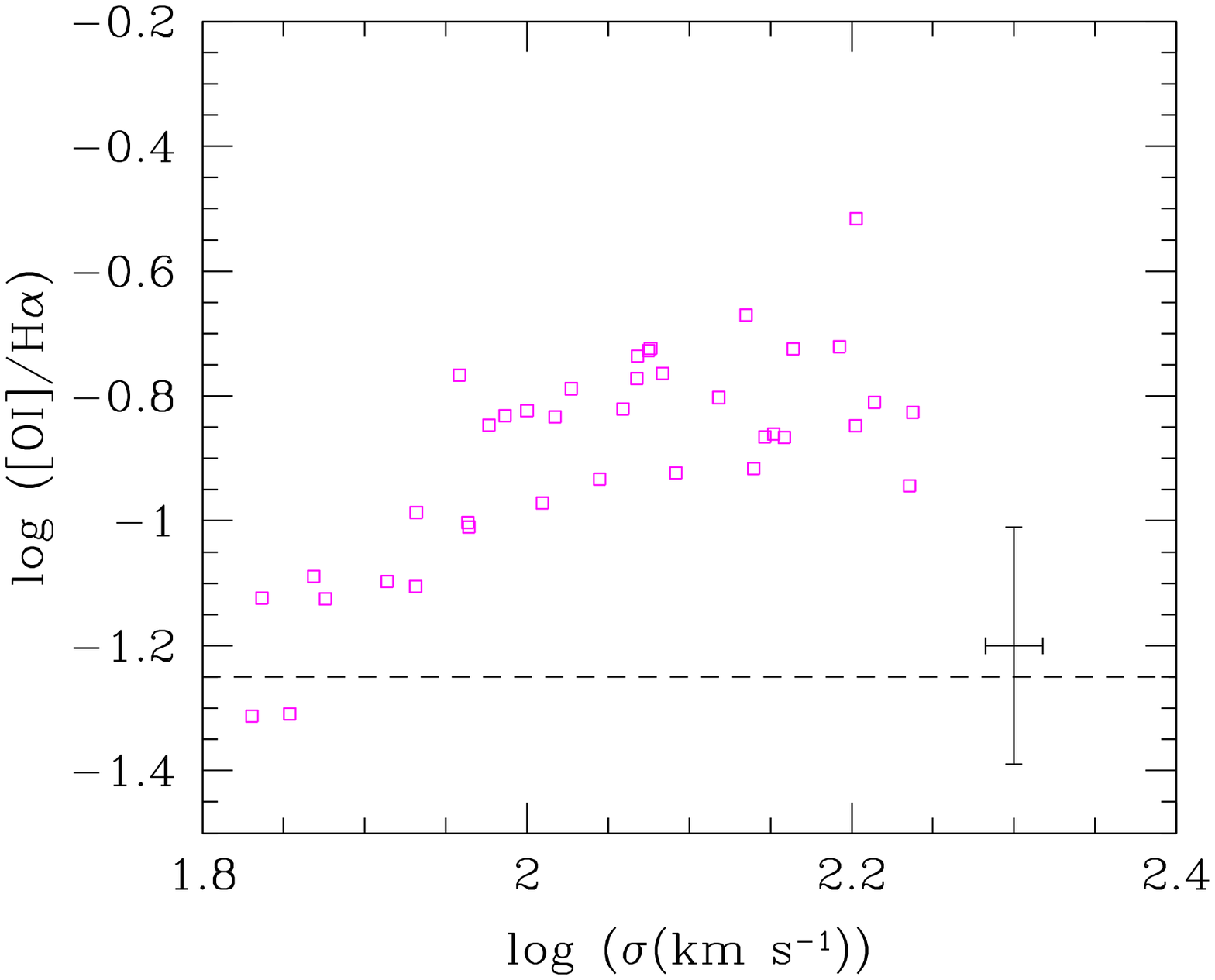}\\
\includegraphics[width=3.6cm,angle=0, clip=,bbllx=30, bblly=275, bburx=560,
bbury=640]{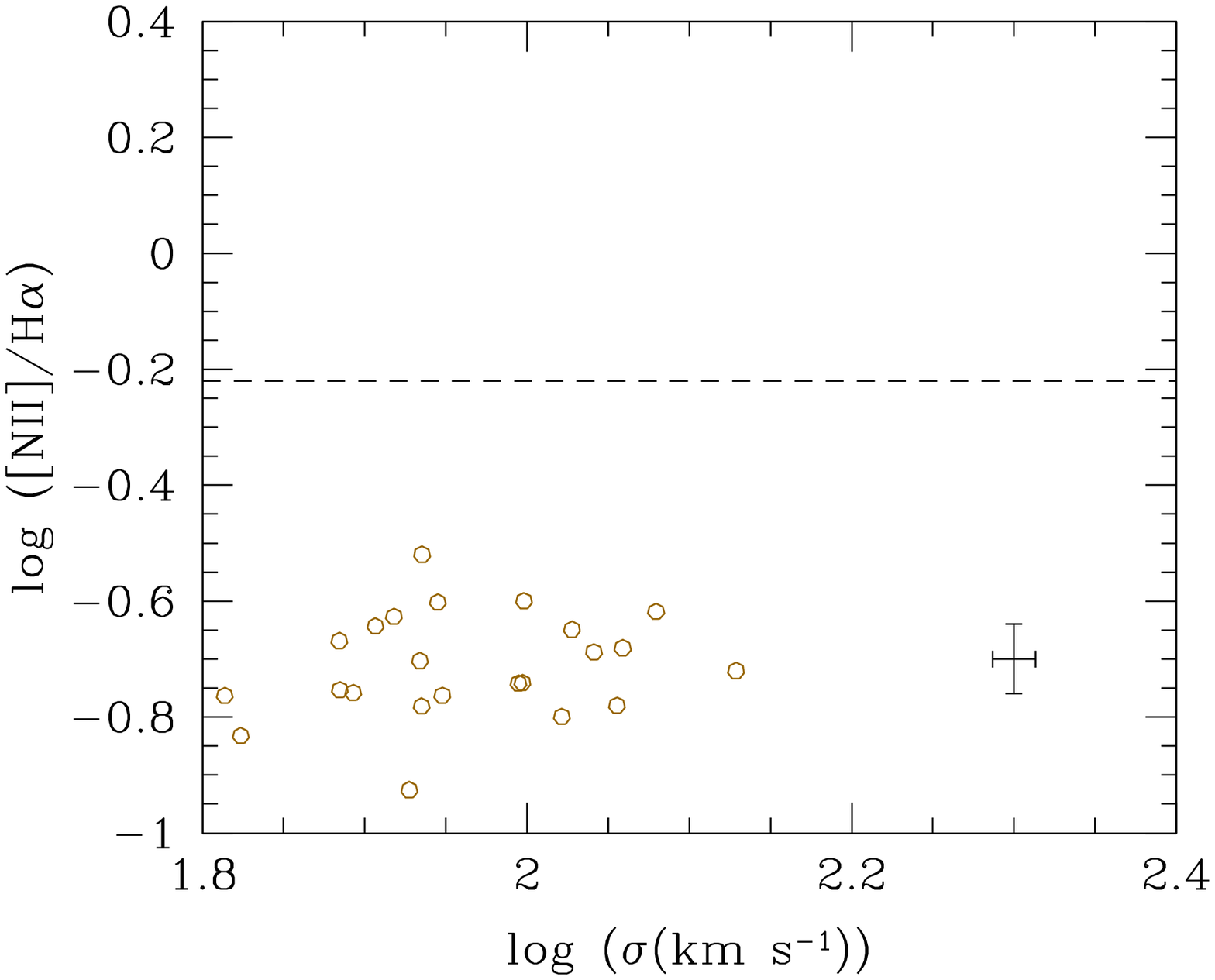} &
\includegraphics[width=3.6cm,angle=0, clip=,bbllx=30, bblly=275, bburx=560,
bbury=640]{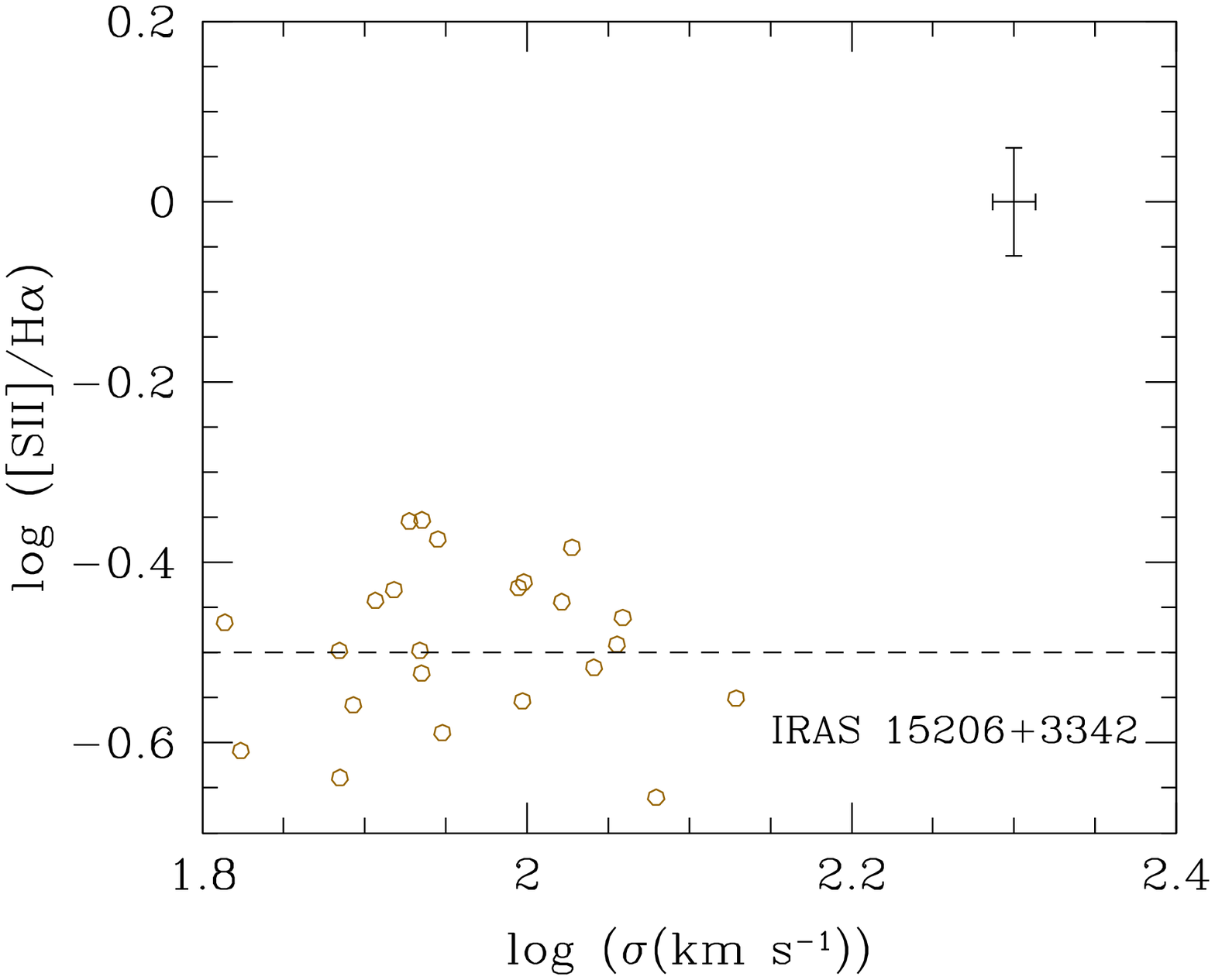} &
\includegraphics[width=3.6cm,angle=0, clip=,bbllx=30, bblly=275, bburx=560,
bbury=640]{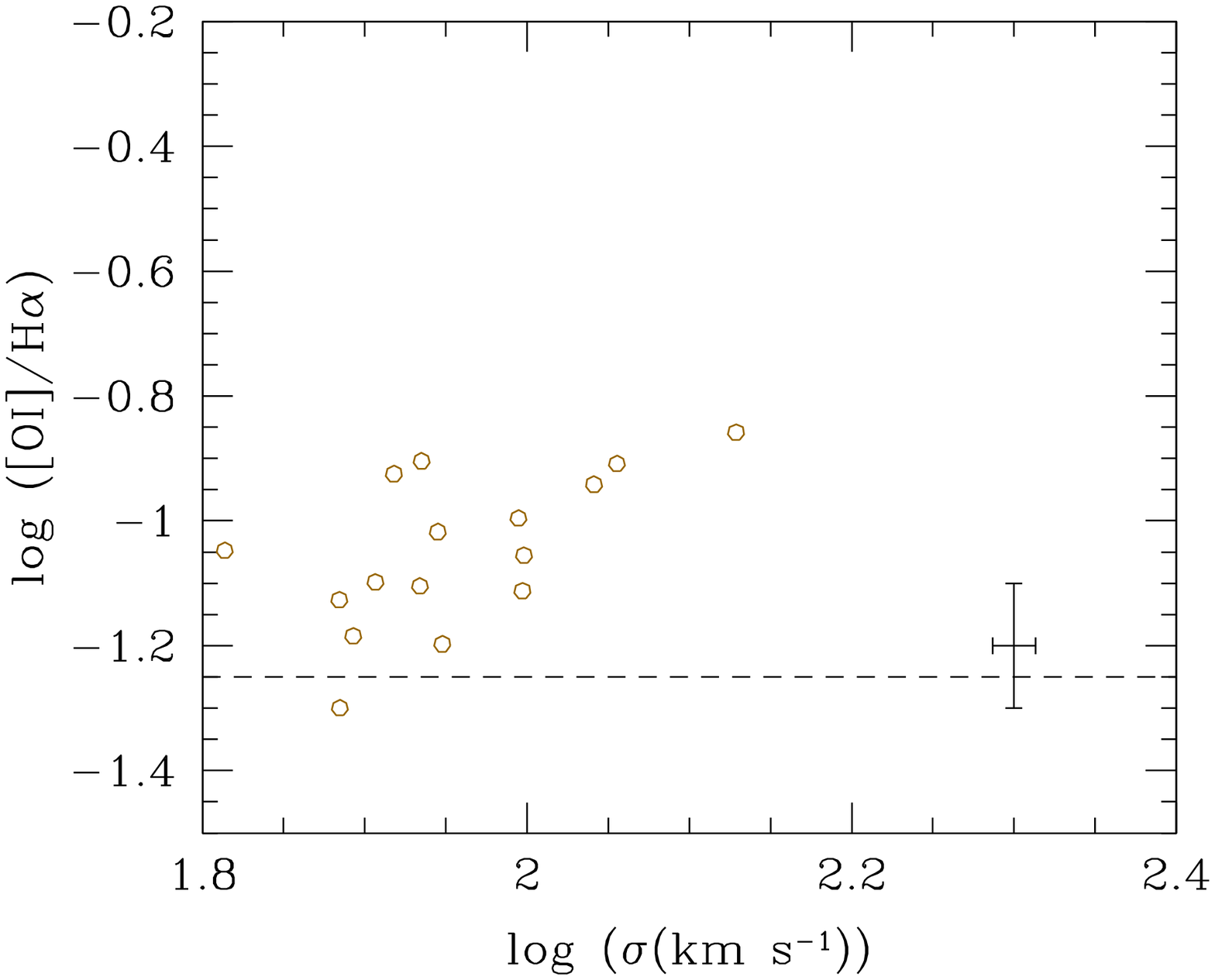}\\ 
\includegraphics[width=3.6cm,angle=0, clip=,bbllx=30, bblly=275, bburx=560,
bbury=640]{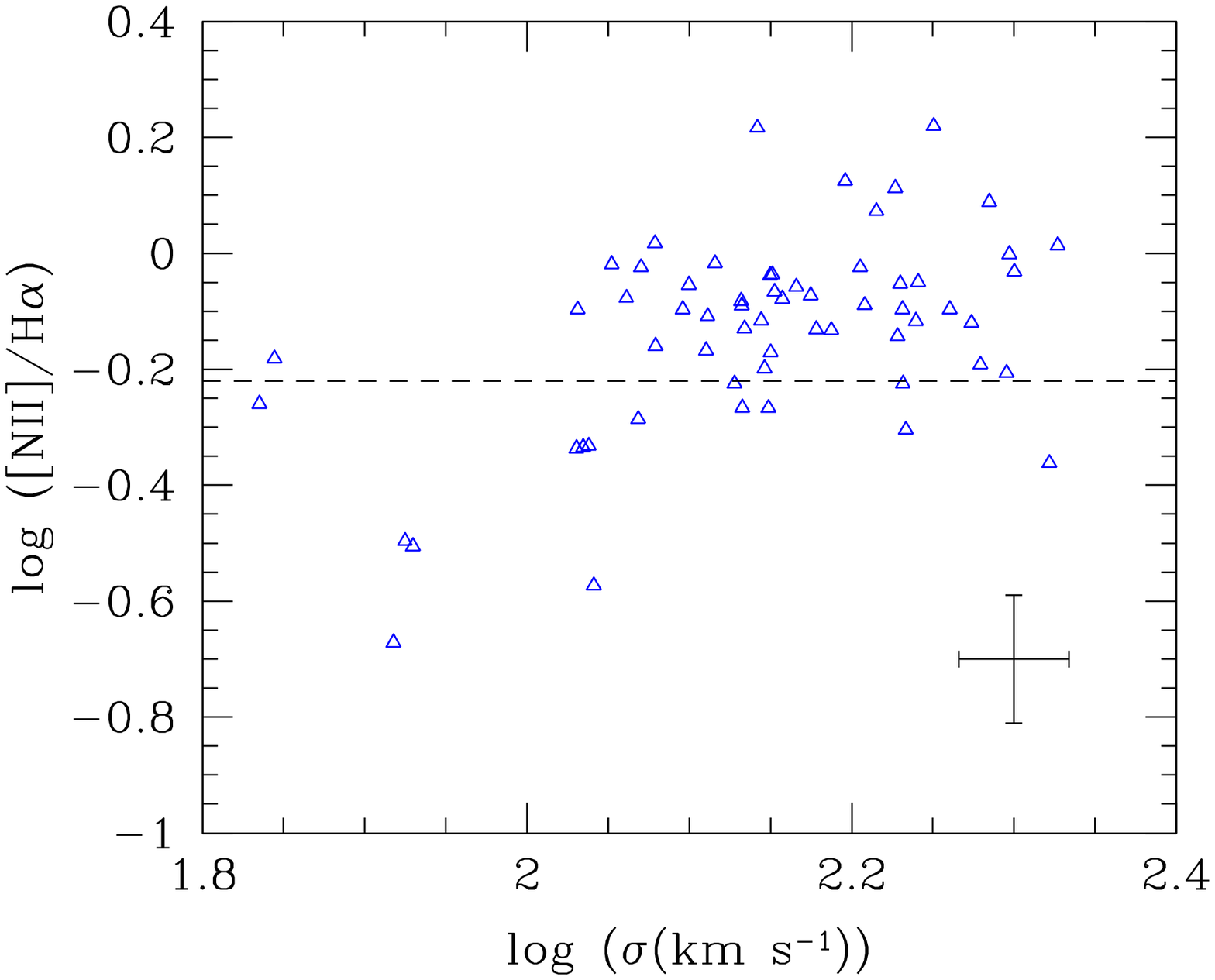} &
\includegraphics[width=3.6cm,angle=0, clip=,bbllx=30, bblly=275, bburx=560,
bbury=640]{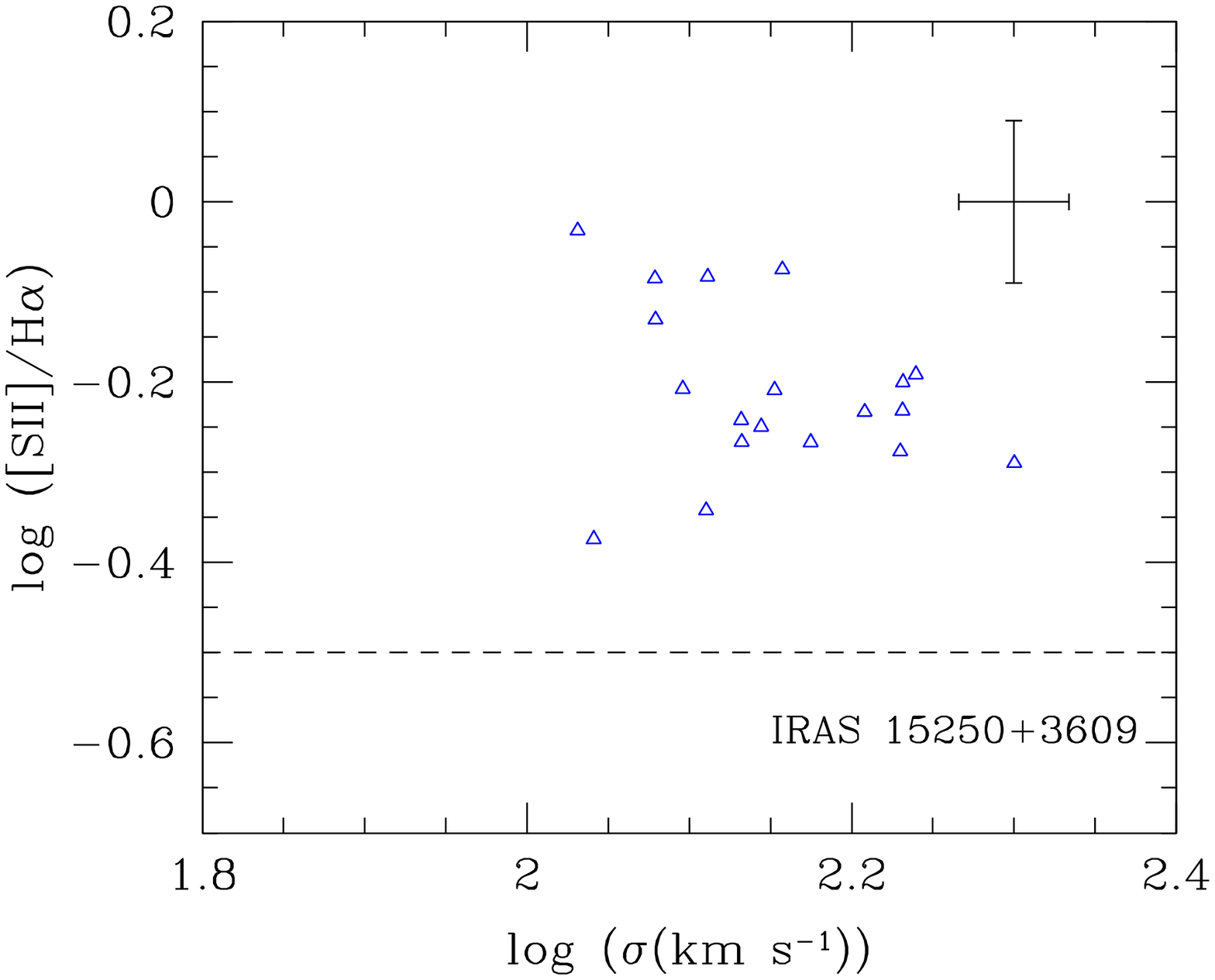} &
\includegraphics[width=3.6cm,angle=0, clip=,bbllx=30, bblly=275, bburx=560, 
bbury=640]{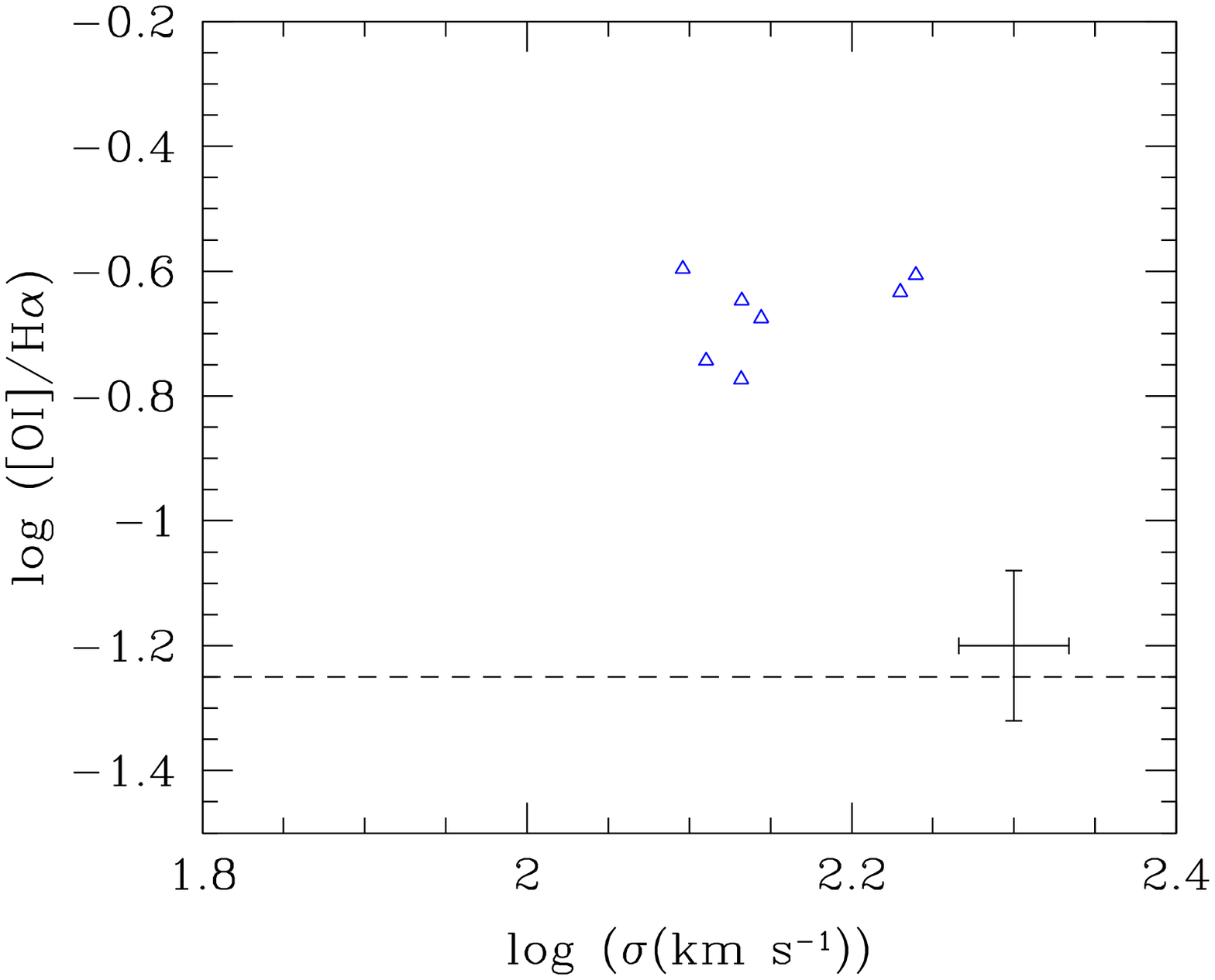}\\ 
\includegraphics[width=3.6cm,angle=0, clip=,bbllx=30, bblly=210, bburx=560,
bbury=640]{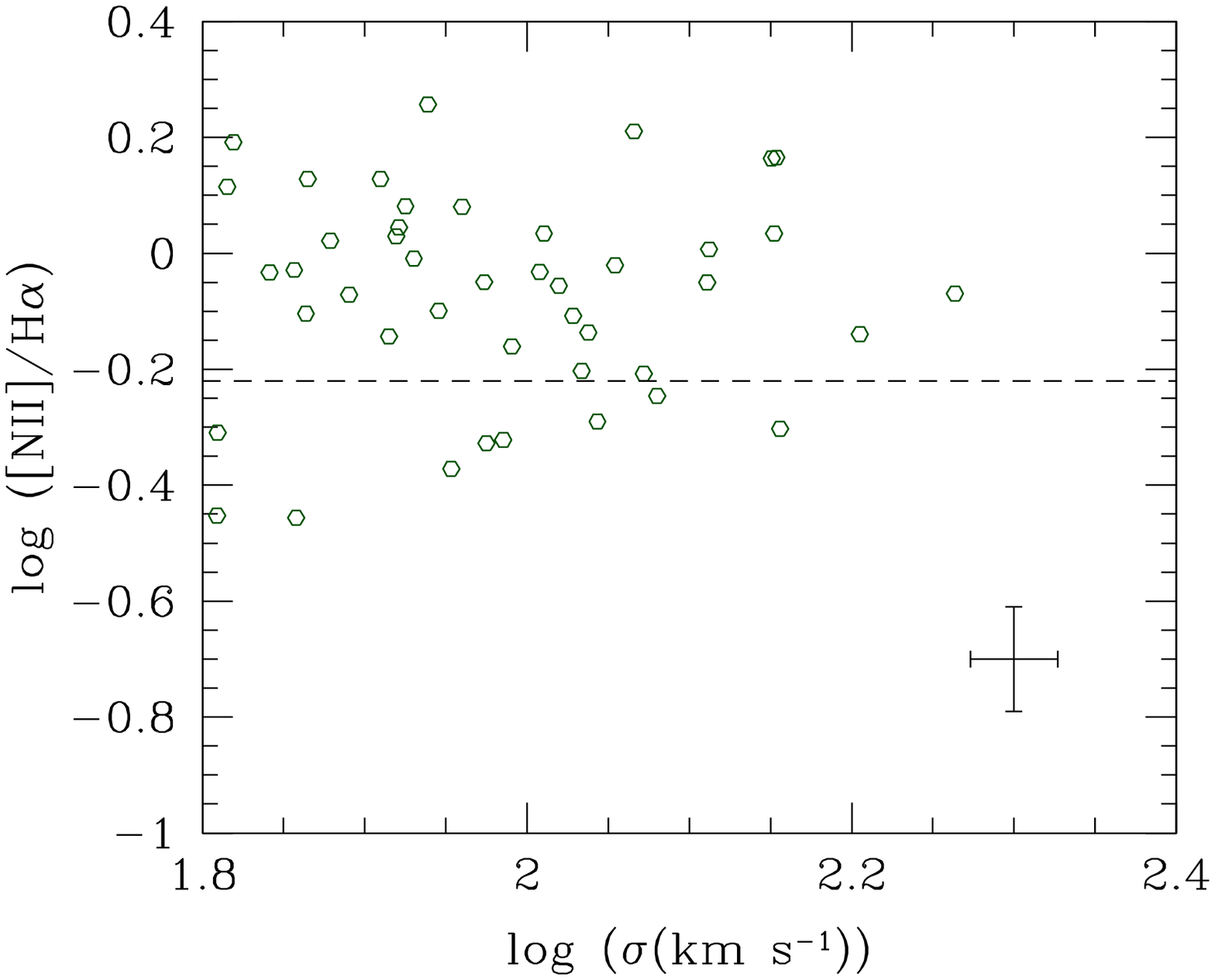} &
\includegraphics[width=3.6cm,angle=0, clip=,bbllx=30, bblly=210, bburx=560,
bbury=640]{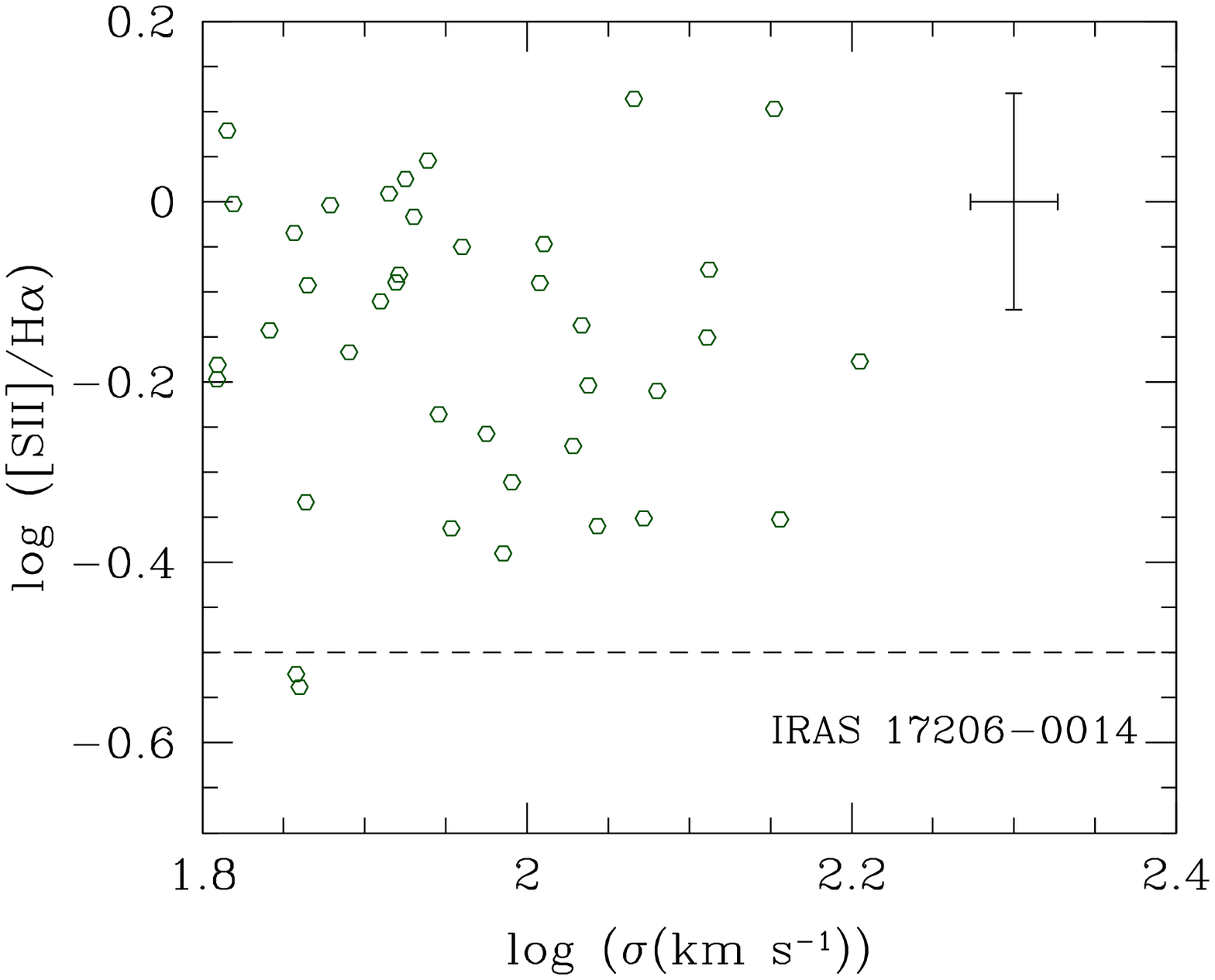} &
\includegraphics[width=3.6cm,angle=0, clip=,bbllx=30, bblly=210, bburx=560,
bbury=640]{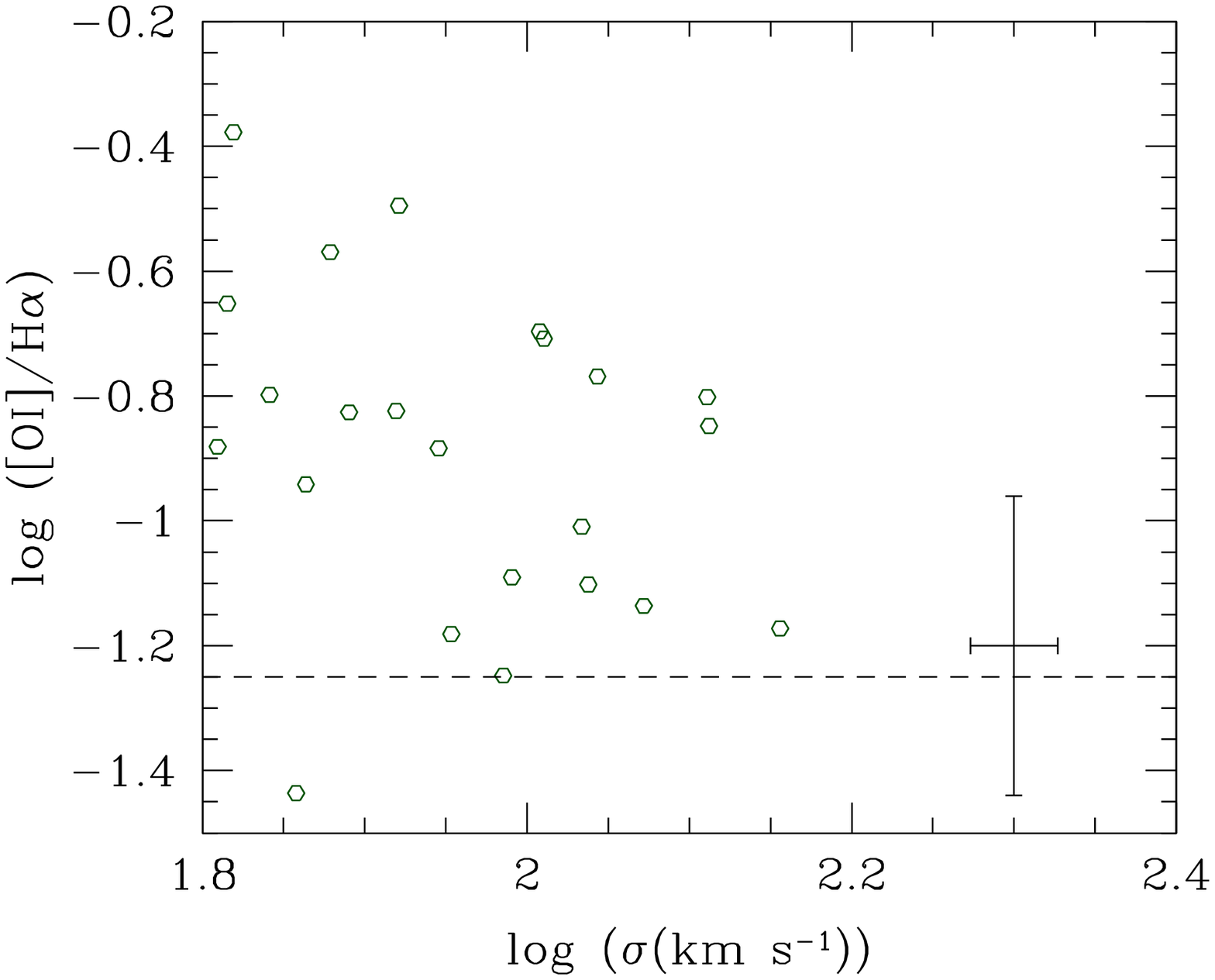}\\ 

\end{tabular}
\protect\caption[Relation between the velocity dipersions and the line
ratios.]
{Relation between the velocity dispersion and 
  \textsc{[N$\;$ii]}$\lambda$6584/H$\alpha$ (left),
  \textsc{[S$\;$ii]}$\lambda\lambda$6717,6731/H$\alpha$ 
  (middle) and \textsc{[O$\;$i]}$\lambda$6300/H$\alpha$ (right). Dashed
  horizontal line marks the frontiers between \textsc{H$\;$ii} region and
  LINER-type ionization. First row shows the combined data for all the systems
  with the exception of IRAS~17206$-$0014. Following rows show data for the
  individual systems. The mean errors for each galaxy are shown in the
  corners of the graphics.
\label{cociydisp}}

\end{figure}
\clearpage

\begin{figure}

\epsscale{1.00}
\plottwo{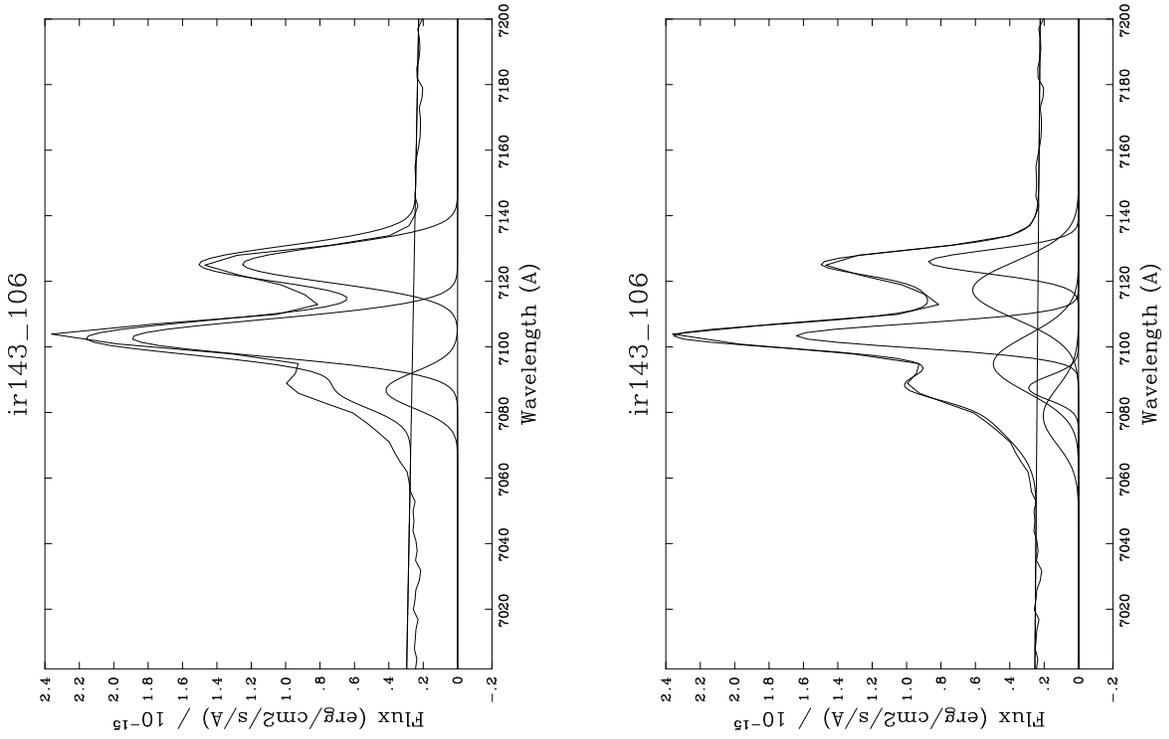}{./f5b.eps}
\caption{Fitting of H$\alpha$+\textsc{[N$\;$ii]}$\lambda\lambda$6548,6584
  lines in IRAS~14348$-$1447 to one or
  two sets of gaussians for the fiber \#106. \label{ajusir143}}
\end{figure}

\clearpage

\begin{deluxetable}{cccccccc}
\tabletypesize{\small}
\rotate
\tablecaption{ULIRGs sample \label{misulirgs}}
\tablewidth{0pt}
\tablehead{
\colhead{Galaxy} & \colhead{$z$} & \colhead{Scale} &
\colhead{$\log (L_{\mathrm{IR}}/L_\odot)$} & \colhead{$f_{12}$\tablenotemark{1}} &
\colhead{$f_{25}$\tablenotemark{1}} & \colhead{$f_{60}$\tablenotemark{1}} &
\colhead{$f_{100}$\tablenotemark{1}}\\
\colhead{} &\colhead{} &\colhead{(kpc arcsec$^{-1}$)} &\colhead{} &\colhead{(Jy)} &\colhead{(Jy)} &\colhead{(Jy)} &\colhead{(Jy)} 
}
\startdata
IRAS~08572+3915   & 0.058\tablenotemark{2} & 1.20 & 12.15 & 0.32 &
1.70 & 7.43 & 4.59 \\ 
IRAS~12112+0305   & 0.073\tablenotemark{2} & 1.52 & 12.30 & 0.11 &
0.51 & 8.50 & 9.98 \\
IRAS~14348$-$1447 & 0.083\tablenotemark{2} & 1.72 & 12.31 & 0.14 &
0.49 & 6.87 & 7.07 \\ 
IRAS~15206+3342   & 0.125\tablenotemark{2} & 2.60 & 12.18 & 0.08 &
0.35 & 1.77 & 1.89 \\
IRAS~15250+3609   & 0.054\tablenotemark{3} & 1.12 & 12.03 & 0.20 &
1.32 & 7.29 & 5.91 \\ 
IRAS~17208$-$0014 & 0.043\tablenotemark{3} & 0.89 & 12.40 & 0.19 &
1.66 & 3.11 & 3.49 \\
\enddata

\tablenotetext{1}{\citet{mos93}}
\tablenotetext{2}{\citet{kim98}}
\tablenotetext{3}{\citet{kim95}}
\end{deluxetable}
\clearpage

\begin{deluxetable}{cccccccc}
\tabletypesize{\small}
\rotate
\tablecaption{Integral Field Observations\label{obs}}
\tablewidth{0pt}
\tablehead{
\colhead{Galaxy} & \colhead{RA} & \colhead{Dec} & \colhead{Spectral Range} &
\colhead{t$_{\mathrm{exp}}$} & \colhead{Air mass} & \colhead{P.A.} & \colhead{Date}\\  
 & \colhead{hh:mm:ss} & \colhead{gr:mm:ss} & \colhead{(\AA)}
 & \colhead{(s)} & & \colhead{($^\mathrm{o}$)} &  
}
\startdata
IRAS~08572+3915   & 09:00:25.4 & +39:03:54.1 & 5200$-$8100 & 
1800$\times$6 & 1.093 & 0.0 & 1998 Apr 01\\
IRAS~12112+0305   & 12:13:46.0 &+02:48:41.0 & 4900$-$7900 &
1800$\times$5 & 1.178 & 0.0 & 1998 Apr 02\\ 
                  & 12:13:46.3 &+02:48:29.7 & 5100$-$8100 &
1500$\times$4 & 1.143 & 180.0 & 2001 Apr 14\\ 
IRAS~14348$-$1447 & 14:37:38.4 &$-$15:00:22.8 & 5200$-$8200 &
1800$\times$4 & 1.438 & 0.0 & 1998 Apr 01\\ 
IRAS~15206+3342   & 15:22:38.0 & +33:31:36.6 & 5000$-$8100 & 
1800$\times$4 & 1.095 & 0.0 & 1998 Apr 03\\
IRAS~15250+3609   & 15:26:59.4 & +35:58:37.6   & 4900$-$7900 &
1800$\times$5 & 1.031 & 0.0 & 1998 Apr 02\\ 
IRAS~17208$-$0014 & 17:23:22.0 & $-$00:17:00.1 & 5000$-$8100 &
1800$\times$4 & 1.243 & 180.0 & 1998 Apr 03\\
\enddata
\end{deluxetable}
\clearpage
\end{document}